\documentclass[a4paper,11pt]{article}
\usepackage{jheppub} 
\usepackage{lineno}
\usepackage{dsfont}

\AddToHook{bfseries}{\boldmath}

\title{Direct simulation Monte Carlo for dark matter production and phase-space evolution}

\author[a,b]{Felix Kahlhoefer}
\author[b]{and Ann Cathrin Wesener}

\affiliation[a]{Institute for Theoretical Particle Physics (TTP),
Karlsruhe Institute of Technology (KIT), 76128 Karlsruhe, Germany}

\affiliation[b]{Institute for Astroparticle Physics (IAP), Karlsruhe Institute of Technology (KIT),
Hermann-von-Helmholtz-Platz 1, 76344 Eggenstein-Leopoldshafen, Germany}

\emailAdd{kahlhoefer@kit.edu}
\emailAdd{wesener@kit.edu}

\abstract{To study the early-universe production of dark matter particles beyond thermal equilibrium often requires solving several coupled Boltzmann equations at the phase-space level. We present a new method to address this challenge using Direct Simulation Monte Carlo (DSMC) based on stochastic interactions of an ensemble of particles. We implement and validate efficient algorithms to describe freeze-out, decays, elastic scattering and number-changing processes within the dark sector. 
These algorithms do not require a discretisation of momentum space on a pre-defined grid and do not rely on approximations such as thermally averaged cross sections.
As illustrative example, we consider a scenario with two-step dark matter production, leading to a bimodal phase-space distribution, and consider the subsequent evolution under elastic and number-changing processes. We show that DSMC is a flexible approach to explore out-of-equilibrium processes and solve Boltzmann equations at the phase-space level in order to predict the abundance of dark matter particles and their effect on structure formation.}


\begin{document}

\noindent{\hfill Preprint: TTP26-043}

\vspace{-1.2em}
\maketitle
\raggedbottom

\section{Introduction}
\label{sec:intro}

The Boltzmann equation is used in many areas of physics to build the bridge from microphysics to macroscopic thermodynamic behaviour. It describes the time evolution of the phase-space distribution function, with the effects of particle motion and interactions encoded in the Liouville and collision operators, respectively.  Since the Boltzmann equation is generally difficult to solve, different areas of physics have developed different strategies tailored to the specific context and available approximations.

In cosmology, for example, one can often assume that distributions are spatially homogeneous and that particle species are in kinetic equilibrium, in which case the collision operator can be integrated to give ordinary differential equations for number and energy densities. This approach is commonly used for example to calculate the relic abundance of dark matter (DM) following departure from chemical equilibrium, the so-called freeze-out mechanism~\cite{Gondolo:1990dk,Griest:1990kh}. However, as the DM particles depart from thermal equilibrium, the approximations underlying the integrated Boltzmann equations become less accurate~\cite{Bringmann:2006mu,Binder:2017rgn,Profumo:2025uvx}. It has been shown that the simple freeze-out calculation can give significantly biased results in scenarios with early kinetic decoupling~\cite{Binder:2017rgn}. 

If DM particles maintain kinetic equilibrium among themselves through self-scatterings, a possible strategy is to include higher moments of the Boltzmann equation~\cite{Binder:2017rgn} leading to a system of coupled differential equations. If self-scattering is inefficient, the most common approach is to return to the unintegrated form of the Boltzmann equation and solve it by discretising phase space, an approach first developed in the context of neutrino decoupling~\cite{Hannestad:1995rs} and then adopted also to the case of DM~\cite{Binder:2021bmg,Hryczuk:2022gay,Chatterjee:2025vdz,DEramo:2025fvy}. This approach however quickly becomes numerically very expensive in spite of various strategies that have been proposed recently to improve its efficiency~\cite{Yoon:2026rce,Spinner:2026ytq,Cervantes:2026fbk}. A particularly challenging situation arises if DM particles never achieve thermal equilibrium, as in the case of freeze-in production~\cite{Hall:2009bx}. In such a scenario the phase-space distribution function may differ significantly from the thermal form and can even be multimodal~\cite{McDonald:2015ljz,Roland:2016gli,Heeck:2017xbu,Du:2021jcj}, making an efficient discretisation of momentum space particularly difficult.

A promising way forward is to consider strategies developed in other branches of physics, which commonly study systems that are spatially inhomogeneous and/or far from thermal equilibrium. In the study of rarefied gas dynamics, for example, a widely used strategy to solve the Boltzmann equation is the so-called Direct Simulation Monte Carlo (DSMC) method, introduced in the 1960s \cite{Bird1963, Bird1970, Bird1994}. This method considers a random sample of particles drawn from the phase-space distribution function and simulates their time evolution by alternating free-streaming and stochastic scattering processes.
It has been applied in a wide range of areas, ranging from space engineering \cite{Pfeiffer2016} to heavy ion collisions \cite{Xu:2004mz, SMASH:2016zqf}.

DSMC methods turn out to be particularly advantageous for dilute systems far away from thermal equilibrium, which contain particles with a wide range of different momenta, because they do not require a parametrised functional form for the phase-space distribution and naturally accommodate even broad ranges of momenta.
Moreover, DSMC methods avoid explicit evaluation of the full collision integral and therefore perform well in models with complicated collision processes involving more than two particles in the initial and/or final state. Finally, DSMC methods provide control over statistical uncertainties, which can be assessed by repeating simulations and reduced proportional to $1/\sqrt{N}$ by increasing the number $N$ of simulation particles. Indeed, several recent studies have shown that DSMC methods can be very useful for cosmology to study neutrino decoupling and derive constraints on physics beyond the Standard Model from Big Bang Nucleosynthesis~\cite{Ovchynnikov:2024rfu,Ihnatenko:2025kew,Akita:2026xwu}. These studies find that DSMC methods often outperform alternative approaches based on discretised momentum space or parametrised distribution functions.

In the present work we investigate possible applications of DSMC in the context of the production of DM particles and their subsequent phase-space evolution. These considerations are particularly important in models where interactions between DM particles and Standard Model (SM) states are very feeble, as for freeze-in production. Since such DM particles are extremely difficult to detect in laboratory experiment, astrophysical and cosmological probes of the matter power spectrum~-- using for example the Lyman-alpha forest, the abundance and dynamics of dwarf galaxies or gravitational lensing of galaxies and clusters~-- provide the most promising strategy to study the properties of DM (see ref.~\cite{Nadler:2026waz} for a recent review). At the same time it becomes very challenging to accurately predict these observables, because thermal equilibrium cannot be assumed at any point in cosmological history and phase-space distributions can differ substantially from the equilibrium form~\cite{Bae:2017dpt,Kamada:2019kpe,Dvorkin:2020xga,Ballesteros:2020adh,DEramo:2020gpr,Egana-Ugrinovic:2021gnu,Decant:2021mhj,Dienes:2021cxp,Xu:2024uas}.

To illustrate different applications of DSMC, we consider a toy model in which DM particles are produced through the decays of a heavy scalar particle that freezes out of thermal equilibrium. We use DSMC to describe the freeze-out beyond kinetic equilibrium, the decays of the decoupled scalars as well as the subsequent interactions between DM particles. We consider both elastic $2\leftrightarrow 2$ scattering, which can redistribute energy between relativistic and non-relativistic particles and thereby change the evolution of the energy density in an expanding universe, as well as $2 \leftrightarrow 3$ number-changing processes, which convert kinetic energy into rest mass and thereby change both the abundance of DM particles and their average momentum~\cite{Carlson:1992fn,Pappadopulo:2016pkp,Cervantes:2024ipg}. Wherever possible, we validate our solutions against alternative ways to solve the Boltzmann equation analytically or numerically. In some cases, however, we obtain new results that cannot be easily obtained using other methods.

We show that the Monte Carlo error decreases as expected when increasing the number of simulation particles and that already a particle number of order $10^5$ is sufficient to predict the relevant observables at the level of precision required by observations. While the present work considers constant matrix elements for simplicity, the algorithms that we consider can be easily adapted to include more complicated cases. In fact, our set-up is well suited for models that require Monte Carlo integration and importance sampling from the final-state phase space. The present study therefore constitutes a first step towards the precision analysis of realistic DM models beyond thermal equilibrium.

The remainder of this work is structured as follows. In section~\ref{sec:set-up} we introduce the Boltzmann equation and the most common strategies for solving it. We furthermore define the toy model considered in the following. Section~\ref{sec: DM DSMC} discusses in detail the DSMC approach, the implemented algorithm and various validation tests. Our main results are presented in section~\ref{sec: results} followed by our conclusions in section~\ref{sec:conclusions}.

\section{General set-up}
\label{sec:set-up}

\subsection{Phase-space evolution}
\label{ch: phase space}

The evolution of the phase-space distribution $f_x(p,t)$ of a particle species $x$ with internal degrees of freedom $g_x$ in a homogeneous and isotropic universe is governed by the Boltzmann equation 
\begin{align}
    \label{eq: Boltzmann}
    \frac{\partial f_x(p,t)}{\partial t}-Hp\frac{\partial f_x(p,t)}{\partial p}=\frac{\hat{C}[f_x]}{E} \; ,
\end{align}
where $H$ denotes the Hubble rate,\footnote{We use the convention that $H =  \sqrt{\frac{g_*}{90}}\pi \frac{T^2}{M_\textrm{pl}}$, where $M_\textrm{pl}=2.4\times 10^{18}\, \textrm{GeV}$ and $g_*$ is the number of relativistic degrees of freedom of the thermal bath, which we take to be constant for simplicity, $g_*=3.36$.} $p$ and $E$ are momentum and energy of the particle $x$ and $\hat{C}[f_x]$ denotes the collision operator, which incorporates all interactions the particle $x$ takes part in. Since the collision operator depends on $f_x$ as well as on the phase-space distributions of all other involved particles, one obtains a system of coupled partial differential equations that is generally hard to solve. 

\paragraph{Moments approach.}
A widely used approach to simplifying the Boltzmann equation is to consider only its lower moments. To take the zeroth moment, one performs the integration $g_x\int\mathrm{d}^3p/(2\pi)^3 $, which results in a differential equation for the number density $n_x=g_x\int\mathrm{d}^3p/(2\pi)^3 f_x$:
\begin{align}
    \label{eq: nBE}
    \dot n_x + 3 H n_x =g_x \int \frac{\mathrm{d}^3p} {(2\pi)^3}\frac{\hat{C}[f_x]}{E} \; .
\end{align}
In order to solve this differential equation, one needs to close the Boltzmann hierarchy by making assumptions regarding the phase-space distributions of the involved particles. In case of kinetic equilibrium of species $x$ with the Standard Model thermal bath, a reasonable assumption is that $f_x\propto \exp(-E/T)$ where $T$ is the thermal bath temperature.\footnote{This approximation implicitly assumes that the particle species under consideration is dilute, i.e.\ that $f_x(p) \ll 1$ for all $p$, such that effects from Pauli blocking or Bose enhancement are negligible.} This assumption is enough in order to solve eq.~\eqref{eq: nBE}. We refer to this approach as the number density Boltzmann equation (\textbf{nBE}).

If the interactions of $x$ with the thermal bath are not strong enough to maintain kinetic equilibrium with it, one needs to weaken the assumption on $f_x$. In case of kinetic equilibrium of $x$ with itself, the species $x$ can be characterised by a separate temperature $T_x$, so that 
$f_x\propto \exp(-E/T_x)$. The temperature is given by
\begin{equation}
\label{eq: effective temperature definition}
    T_x= \bigg\langle \frac{p^2}{3 E} \bigg\rangle_x = \frac{g_x}{n_x}\int \frac{\mathrm{d}^3p}{(2\pi)^3} \frac{p^2}{3 E}\, f_x \; ,
\end{equation}
where the brackets $\langle \rangle_x$ denote the expectation value with respect to the normalised momentum distribution $f_x(p) / n_x$. An equation for the evolution of $T_x$ can then be obtained by performing the corresponding integration over eq.~\eqref{eq: Boltzmann}, which results in
\begin{align}
    \label{eq: cBE}
    3\dot n_x T_x + 3n_x \dot T_x + 15Hn_xT_x -g_xH \int \frac{\mathrm{d}^3p}{(2\pi)^3}\frac{p^4}{E^3}f_x  =g_x \int \frac{\mathrm{d}^3p}{(2\pi)^3}\frac{p^2}{E}\frac{\hat{C}[f_x]}{E} \; .
\end{align}
Eqs.~\eqref{eq: nBE} and \eqref{eq: cBE} form a system of coupled differential equations, which we refer to as the coupled Boltzmann equations (\textbf{cBE}). 

\paragraph{Full Boltzmann equation.}
As soon as the shape of the phase-space distribution differs from the kinetic equilibrium form, the cBE approach is no longer sufficient and one needs to solve the full Boltzmann equation (\textbf{fBE}) in eq.~\eqref{eq: Boltzmann} at the phase-space level. Prominent examples in which a non-trivial phase-space distribution can arise include for example very momentum-selective interactions (e.g. resonances), which deplete or populate only parts of the phase space \cite{DAgnolo:2017dbv, Binder:2021bmg} or scenarios where the production of the species $x$ happens in consecutive steps populating different phase-space regions at different times \cite{Merle:2015oja, Konig:2016dzg, Heeck:2017xbu, Dienes:2020bmn, Benso:2025vgm}. In the following, we will consider the latter scenario and introduce a novel approach for solving the fBE.

\subsection{Toy model}
\label{subsec toy model}
In our set-up, we consider three different particle species: $\psi, \phi$ and $\rho$. For simplicity, we assume that these particles are real scalars and that all their interactions conserve CP symmetry. The particle $\psi$ is assumed to be fully thermalised with the SM thermal bath of temperature $T$ at all relevant times and sufficiently light that its mass is irrelevant for our calculations. It can be thought of as a simplified representation of any of the light SM fermions. The main simplification achieved by assuming $\psi$ to be a real scalar is that we can take the matrix element for the process $\psi \psi \to \phi \phi$ as a constant independent of momentum and scattering angle, which speeds up the Monte Carlo simulations introduced below.

The second particle, $\phi$, on the other hand, has a mass $m_\phi$ that is non-negligible compared to the temperature. It interacts with $\psi$ exclusively through the annihilation process $\psi\psi \leftrightarrow \phi\phi$, which we assume to be initially strong enough so that $\phi$ is in thermal equilibrium with $\psi$ and then undergoes Boltzmann suppression and thermal freeze-out. We neglect all other interactions between $\psi$ and $\phi$, such as scattering processes $\psi \phi \leftrightarrow \psi \phi$ or (inverse) decays $\phi \leftrightarrow \psi \psi$.

\begin{figure}
    \centering
    \includegraphics[width=\linewidth]{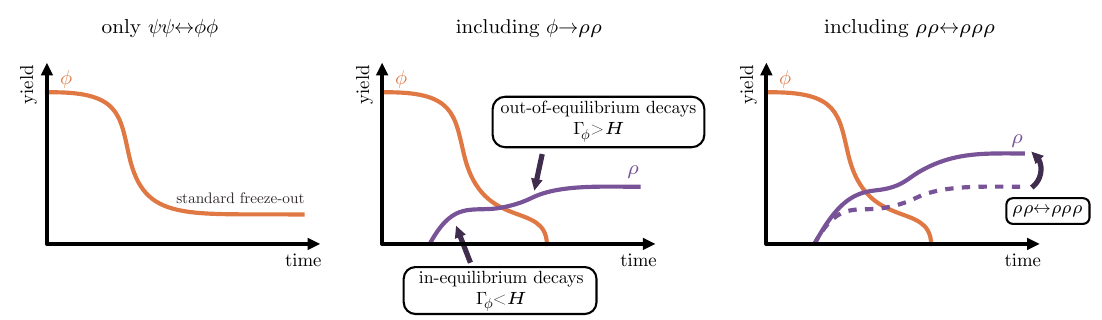}
    \caption{Evolution of the yield including only $\psi\psi \leftrightarrow \phi \phi$ (left), $\phi\to \rho\rho$ (middle) and $\rho\rho\leftrightarrow \rho \rho\rho$ (right).}
    \label{fig: schematic abundance}
    \includegraphics[width=\linewidth]{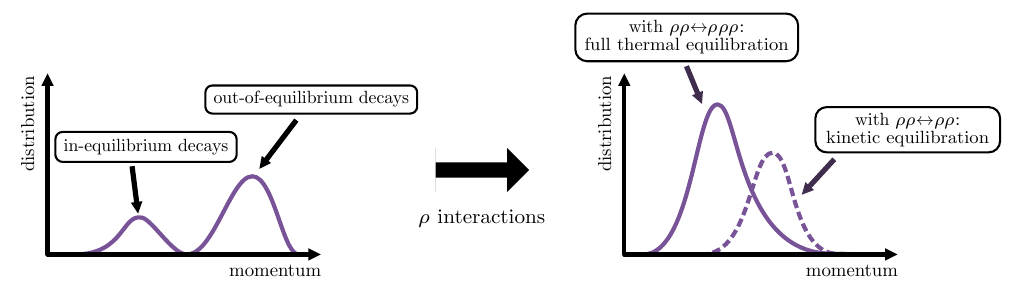}
    \caption{Phase-space distribution of $\rho$ particles after all interactions cease without (left) and with (right) $\rho$ self-interactions.}
    \label{fig: schematic phase space}
\end{figure}

A coupling between $\phi$ and $\rho$ allows for the decay $\phi \to \rho\rho$. Here we assume that $m_\rho \ll m_\phi$ and that the decays are slow enough that the $\rho$ particles never thermalise with the $\psi$ particles. The decays of $\phi$ then happen in two stages: The first stage (in-equilibrium decays) occurs while the $\phi$ particles are still abundant and in thermal equilibrium with the bath, but the decay width is small compared to the Hubble rate. The second stage occurs after the $\phi$ particles have frozen out, once the Hubble rate has decreased enough to become comparable with the decay width. We show a schematic sketch of the abundance evolution in figure \ref{fig: schematic abundance}.
This scenario has been intensely studied for example in ref.~\cite{Du:2021jcj} and is known to produce bimodal phase-space distributions for the $\rho$ particles. This is because in-equilibrium decays produce $\rho$ particles with momentum comparable to the thermal bath, while the out-of-equilibrium decays produce particles with momentum $p \approx m_\phi / 2 \gg T$ as illustrated in the left panel of figure~\ref{fig: schematic phase space}.

While the initial phase-space distribution of $\rho$ particles arising from $\phi$ decays has been studied in great detail in the literature, the focus of our study is on the subsequent evolution of this distribution under self-interactions of $\rho$ particles. We consider not only elastic $2\rho\leftrightarrow 2\rho$ collisions, but include also number-changing interactions of the form $2\rho \leftrightarrow 3\rho$. As we will see below, the latter type of process can lead to large effects because it can convert the large kinetic energy of $\rho$ particles produced in out-of-equilibrium decays into additional particles with less kinetic energy, as shown in the right panel of figure~\ref{fig: schematic phase space}.

\section{Dark Matter Direct Simulation Monte Carlo}
\label{sec: DM DSMC}
The standard approach for solving the unintegrated Boltzmann equation in DM phenomenology goes via discretizing the phase space. This approach has been shown to work reliably, but it has some disadvantages. For example, knowledge of the relevant phase-space areas is needed beforehand to decide on the discretised momentum bins. Furthermore, a large separation of relevant momentum scales necessitates a large momentum lattice which becomes computationally unfeasible. Instead of discretizing the phase space, we present a novel approach based on performing a Monte-Carlo simulation of the involved particles and interactions. We will use the DSMC method and extend it to $2\leftrightarrow3$ collisions. 

\subsection{Overview of DSMC}

\begin{figure}[t]
    \centering
    \includegraphics[width=0.9\linewidth]{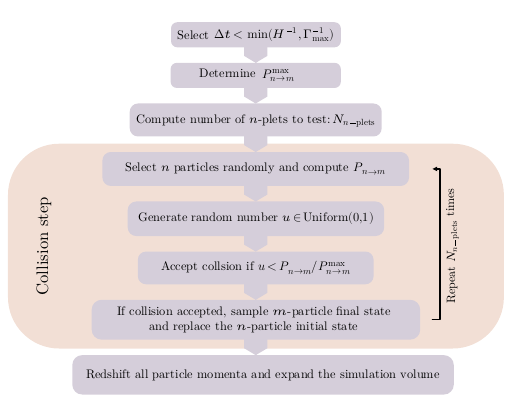}
    \caption{Flowchart of the DSMC scheme for $n\to m$ scattering. For simplicity, we illustrate here the case of just one type of interaction per timestep. The implementation with multiple interactions is described below.}
    \label{fig:DSMC_sketch}
\end{figure}

DSMC solves the full Boltzmann equation by simulating $N$ particles and their interactions in a volume $V$. Since we consider a homogeneous and isotropic universe, we only need to know the energies of the particles in the simulation volume. During each time step $\Delta t$, a collision step is performed, followed by a transport step, see figure~\ref{fig:DSMC_sketch}. During the collision step, particles are randomly selected and assigned with random isotropic momentum directions, and collisions between selected particles are sampled as described below in more detail. During the transport step, the momenta of the particles and the temperature of the thermal bath are redshifted and the simulation volume is expanded. The size of the time step $\Delta t$ is chosen such that $\Delta t \ll \Gamma^{-1}$, where $\Gamma$ denotes all relevant rates, ranging from the Hubble rate to collision and decay rates. 

The collision step is performed using the so-called No-time-counter (NTC) algorithm \cite{Bird1994, ROOHI20161, stefanov2019}. While the literature describes the NTC algorithm only for $2\to 2$ collisions, we extend it here to $n\to m$ collisions. Consider $n$ particles $i_1,\dots, i_n$ randomly selected from the simulation volume. Their probability to perform an $n\to m$ collision is denoted as $P_{i_1,\dots, i_n}$. The brute-force way of performing the collision step would be to go through all $\binom{N}{n}$ $n$-plets,\footnote{This is the number of unordered $n$-plets for identical initial particles. For non-identical initial particles, the number has to be adapted accordingly.} generate a random number $u_{i_1,\dots, i_n}\in \textrm{Uniform}(0,1)$ for each $n$-plet and accept the collision if $u_{i_1,\dots i_n}<P_{i_1,\dots, i_n}$. In general, however, the collision probability may be very small, making this approach highly inefficient. Instead, in the NTC approach, at the beginning of each time step, the maximal possible interaction probability $P_n^\textrm{max}$ among all $n$-plets is determined. Then, 
$N_{n-\textrm{plets}}= \binom{N}{n} \cdot P_{n}^\textrm{max}$ $n$-plets are selected. In practice, since $N_{n-\textrm{plets}}$ is not an integer number, we choose the number of $n$-plets from a Poisson distribution with mean $N_{n-\textrm{plets}}$. For each $n$-plet a collision is accepted if the random number satisfies
\begin{equation}
    u_{i_1,\dots, i_n}< \frac{P_{i_1,\dots, i_n}}{P_n^\textrm{max}} \; .
\end{equation} 

The interaction probability can be determined using the stochastic collision criterion \cite{DANIELEWICZ1991712, Cassing:2001ds, Xu:2004mz, Staudenmaier:2021lrg}, which we review following the discussion in ref.~\cite{Xu:2004mz}.
We first only consider $2\to 2$ scattering, the generalisation to $n\to m$ scattering follows along similar lines. The collision operator for $2\to 2$ scattering is given by 
\begin{align}
\nonumber
    \frac{\hat{C}[f_1]}{E_1}= 
    \frac{1}{2E_1}
\int \mathrm{d}\pi_2
\,\kappa
\int \mathrm{d}\pi_1' \mathrm{d}\pi_2'
\, f_1' f_2'
\, \left| \mathcal{M}_{1'2' \to 12} \right|^2(2\pi)^4
\delta^{(4)}\!\left(p_1' + p_2' - p_1 - p_2\right)
\qquad\\
-
\frac{1}{2E_1}
\int \mathrm{d}\pi_2
\,\kappa
\int \mathrm{d}\pi_1' \mathrm{d}\pi_2'
\, f_1 f_2
\, \left| \mathcal{M}_{12 \to 1'2'} \right|^2
(2\pi)^4
\delta^{(4)}\!\left(p_1 + p_2 - p_1' - p_2'\right),
\end{align}
where $\kappa$ is a symmetry factor accounting for identical particles in the final state ($\kappa=1/2$ for identical final state particles) and $\mathrm{d}\pi_i=\frac{\mathrm{d}^3p_i}{(2\pi)^32E_i}$ is the Lorentz-invariant differential phase-space. 
The collision probability for $1 2 \rightarrow 1'  2'$ when selecting two particles from an ensemble of particles in a volume $V$ can be easily obtained by reinterpreting the Boltzmann equation in eq.~\eqref{eq: Boltzmann} for vanishing Hubble expansion: The remaining parts of the Boltzmann equation describe how the phase-space distribution of particle 1 changes solely due to scattering. The phase-space distribution is related to the number of particles $N_i$ in a given unit phase-space box $\Delta x^3 \Delta p^3$ around phase-space coordinate $(x,p)$ through 
\begin{align}
    f_i(x,p)=\frac{N_i}{\frac{1}{(2\pi)^3}\Delta x^3 \Delta p^3}
\end{align}
assuming no internal degrees of freedom. 

The change in the number of particles 1 corresponds to the number of collisions: $\Delta N_1 = N_\text{coll}$. The number of collisions of $N_1$ particles of type 1 in momentum range $(\bf{p}_1, \bf{p}_1+\Delta p_1)$ and $N_2$ particles of type 2 in the momentum range $(\bf{p_2}, \bf{p_2}+\Delta p_2)$ in a volume $\Delta^3x$ and time interval $\Delta t$ is given by 
\begin{align}
\nonumber
    \frac{\partial f(p_1,t)}{\partial t}\rightarrow \frac{N_\textrm{coll}}{\Delta t\frac{1}{(2\pi)^3}\Delta x^3 \Delta p_1^3 }=\frac{1}{2E_1}\frac{\Delta p_2^3}{(2\pi)^32E_2}\kappa f_1f_2
    \int & \mathrm{d}\pi_1' \mathrm{d}\pi_2'
\left| \mathcal{M}_{12 \to 1'2'} \right|^2 \\
& \times (2\pi)^4
\delta^{(4)}\!\left(p_1 + p_2 - p_1' - p_2'\right).
\end{align}
The collision probability for a randomly drawn pair of particles in the momentum ranges $(\bf{p}_1, \bf{p}_1+\Delta p_1)$ and  $(\bf{p_2}, \bf{p_2}+\Delta p_2)$ is therefore given by
\begin{align}
\nonumber
    P_{2\rightarrow 2}&=\frac{N_\textrm{coll}}{N_1 N_2}\\\nonumber
    &=\frac{1}{2E_1}\frac{1}{2E_2}\frac{\Delta t}{\Delta x^3}\kappa
    \int \mathrm{d}\pi_1' \mathrm{d}\pi_2'\, \left| \mathcal{M}_{12 \to 1'2'} \right|^2
(2\pi)^4
\delta^{(4)}\!\left(p_1 + p_2 - p_1' - p_2'\right)\\&=\sigma_{2\rightarrow 2}v\frac{\Delta t}{\Delta x^3} \; ,
\end{align}
where we used the definition of the cross section
\begin{align}
    \sigma_{2\rightarrow 2}v=\frac{1}{2E_1}\frac{1}{2E_2}\kappa
    \int \mathrm{d}\pi_1' \mathrm{d}\pi_2'\, \left| \mathcal{M}_{12 \to 1'2'} \right|^2
(2\pi)^4
\delta^{(4)}\!\left(p_1 + p_2 - p_1' - p_2'\right)
\end{align}
with the relative velocity $v$ of the two incoming particles in the cosmic frame. 

The expression for $ P_{2\rightarrow 2}$ corresponds to the standard geometric interpretation of the cross section, where $\sigma_{2\rightarrow 2}v\Delta t$ is the "collision area" swept by the two incoming particles during a time step $\Delta t$. The advantage of this derivation of the collision probability is that it can easily be generalised to $n\to m$ scattering by using the corresponding collision integrals. In the present work we restrict ourselves to $2\leftrightarrow 2$ and $2\leftrightarrow3$ scatterings. The collision probabilities for $2\rho\to 3\rho$ and $3\rho\to 2\rho$ scatterings are derived in appendix~\ref{app: 2 to 3} and are given by 
\begin{align}
    P_{2\rightarrow3}=\sigma_{2\rightarrow 3}v\frac{\Delta t}{\Delta x^3}
\end{align}
with 
\begin{align}
    \sigma_{2\rightarrow 3}v=\frac{1}{2E_1}\frac{1}{2E_2}\frac{1}{3!}
    \int \mathrm{d}\pi_1' \mathrm{d}\pi_2' \mathrm{d}\pi_3'\, \left| \mathcal{M}_{12 \to 1'2'3'} \right|^2 
(2\pi)^4
\delta^{(4)}\!\left(p_1 + p_2 - p_1' - p_2'-p_3'\right)
\end{align}
and 
\begin{align}
\label{eq: scattering prob 3 to 2}
    P_{3\rightarrow2}=\frac{1}{8E_1E_2E_3}\frac{\Delta t}{(\Delta x^3)^2}I_{32} \; ,
\end{align}
where we defined the Lorentz-invariant integral expression
\begin{align}
    I_{32}=\frac{1}{2!}
    \int \mathrm{d}\pi_1' \mathrm{d}\pi_2'
\, \left| \mathcal{M}_{123 \to 1'2'} \right|^2
(2\pi)^4
\delta^{(4)}\!\left(p_1 + p_2 +p_3- p_1' - p_2'\right) \; .
\end{align}

\subsection{Implemented algorithm}

Determining the maximum possible interaction probability for the particles in the simulation volume is a crucial step in the NTC scheme. The stricter the upper bound, the higher the acceptance rate of collisions, which heavily influences the runtime. To optimise the runtime, methods adapted to the specific particles and interactions are needed. For the model we consider as an example, we will describe these methods in the following.

\paragraph{$\psi\psi \to \phi \phi$ production:} Since we assume that $\psi$ is in thermal equilibrium with the SM bath, we can sample collision pairs from the thermal probability density function
\begin{equation}
    g_\textrm{eq}(p)=\frac{p^2}{2 T^3} \exp(-p/T) 
\end{equation}
for temperature $T$ and negligible mass $m_\psi$. To be consistent with the implemented collision scheme where we do not include Bose enhancement or Pauli blocking, we neglect here quantum statistics factors in the momentum distribution.  Due to the assumed hierarchy $m_\phi \gg m_\psi$, only high-energy $\psi$ particles can annihilate into $\phi$ particles simply due to kinematics. Sampling random pairs from the $\psi$ bath and testing them for collisions according to the NTC algorithm thus results in a huge overhead of rejected collisions. 

Instead of sampling pairs directly from the thermal distribution, one can introduce sequential sampling, which takes into account the energy threshold for successful annihilation: An annihilation is only possible if the maximal possible center-of-mass (CM) energy of the selected pair $s_\textrm{max}=4p_1p_2$ is larger than $4m_\phi^2$. 
Sampling the momenta of the two incoming $\psi$ particles, $p_1$ and $p_2$, independently of the thermal distribution and accepting a pair only if $p_1p_2>m_\phi^2$ corresponds to sampling from the joint distribution $g(p_1, p_2)= g_\textrm{eq}(p_1)g_\textrm{eq}(p_2)\theta(p_1p_2-m_\phi^2)/\mathcal{N}$, where $\mathcal{N}$ is given by the fraction of pairs that satisfy the threshold condition
\begin{align}
    \mathcal{N}= \int \mathrm{d}p_1\mathrm{d}p_2\,g_\textrm{eq}(p_1)g_\textrm{eq}(p_2)\theta(p_1p_2-m_\phi^2) = \frac{1}{2}\left(\frac{m_\phi}{T}\right) ^3\left(2+\frac{m_\phi^2}{T^2}\right)K_3\left(2\frac{m_\phi}{T}\right) \; ,
\end{align}
where $K_i$ denotes the modified Bessel function of the second kind of order $i$.
This is equivalent to sampling $p_1$ from the marginal distribution 
\begin{align}
\nonumber
    g(p_1)&= \int \mathrm{d}p_2\, g(p_1, p_2)= \frac{1}{\mathcal{N}}g_\textrm{eq}(p_1)\int_{m_\phi^2/p_1}^\infty \mathrm{d}p_2 \, g_\textrm{eq}(p_2) \\
    &\propto (m^4_\phi + 2 m_\phi^2T p_1 +2 T^2 p_1^2) \exp\left(-\frac{m^2}{Tp_1}-\frac{p_1}{T}\right)
\end{align}
and then sampling $p_2$ from the conditional distribution  
    \begin{align}
    \nonumber
        g(p_2|p_1)&=\frac{g(p_1,p_2)}{g(p_1)} = \frac{g_\textrm{eq}(p_2)\theta(p_1p_2-m_\phi^2)}{\int_{m_\phi^2/p_1}^\infty \mathrm{d}p_2 \, g_\textrm{eq}(p_2) }\\
        &\propto  p_2^2\exp\left(-\frac{p_2}{T}\right)\theta(p_1p_2-m_\phi^2) \; .
    \end{align}
The sampling from the marginal and conditional distributions is realised using inverse transform sampling with pre-tabulated inverse cumulative distribution functions.
With this sampling procedure, the number of tested $\psi$ pairs has to be reduced by a factor $\mathcal{N}$ to account for the increased acceptance rate. Instead of testing $N_{2-\textrm{plets}}= \binom{N}{2}P^\textrm{max}_{\psi\psi \to \phi \phi}$ random pairs, we thus consider $N_{2-\textrm{plets}}= \mathcal{N} \binom{N}{2}P^\textrm{max}_{\psi\psi \to \phi \phi} $ pre-selected pairs. 
    
Finally, to obtain $P^\textrm{max}_{\psi\psi \to \phi \phi}$, we need an upper bound on the interaction cross section for  $\psi\psi \to \phi \phi$, which is given by 
\begin{align}
\nonumber
        (\sigma v)_{\psi\psi \to \phi \phi} &= \frac{1}{4E_1E_2}\frac{1}{2!}
    \int \mathrm{d}\pi_1' \mathrm{d}\pi_2'\, \left| \mathcal{M}_{\psi\psi \to \phi \phi} \right|^2
(2\pi)^4
\delta^{(4)}\!\left(p_1 + p_2 - p_1' - p_2'\right)\\ &= \frac{ \left| \mathcal{M}_{\psi\psi \to \phi \phi} \right|^2}{64\pi E_1E_2} \sqrt{1-\frac{4m_\phi^2}{s}}
\leq \frac{ \left| \mathcal{M}_{\psi\psi \to \phi \phi} \right|^2 }{64\pi E_1E_2} \sqrt{1-\frac{m_\phi^2}{E_1 E_2}}
\leq \frac{\left| \mathcal{M}_{\psi\psi \to \phi \phi} \right|^2 }{ 96\sqrt{3}\pi m_\phi^2}
\end{align}
where $s$ is the CM energy of the $\psi$ particles. In the first inequality we have used that $s \leq 4 E_1 E_2$ and in the second inequality that $x \sqrt{1 - x} \leq 2/(3 \sqrt{3})$. In our toy model, the matrix element is taken to be constant.

\paragraph{$\phi \phi \to \psi \psi$ annihilation:} In this case there is no kinematic threshold and the annihilation process is allowed for all pairs of $\phi$ particles. The cross-section is given by 
    \begin{align}
        (\sigma v)_{\phi \phi \to \psi\psi} =  \frac{\left| \mathcal{M}_{\psi\psi \to \phi \phi} \right|^2}{64\pi E_1E_2} \; .
    \end{align}
    An upper bound on the cross section for the particles in the simulation is given by 
    \begin{align}
         (\sigma v)^\textrm{max}_{\phi \phi \to \psi\psi}= \frac{ \left| \mathcal{M}_{\psi\psi \to \phi \phi} \right|^2}{64\pi E^\textrm{min}_1E^\textrm{min}_2} 
    \end{align}
    where $E^\textrm{min}_1$ and $E^\textrm{min}_2$ are the lowest and second-lowest energies of the $\phi$ particles in the cell.

\paragraph{$\phi \to \rho \rho$ decays:} For a rest-frame decay rate $\Gamma_\textrm{dec}$, the decay probability of a particle with energy $E_\phi$ in a time interval $\Delta t\ll\Gamma_\textrm{dec}^{-1}$ is given by 
\begin{align}
    P_\textrm{dec} = \frac{m_\phi}{E_\phi}\Gamma_\textrm{dec}\Delta t \; .
\end{align}
In principle, these decays could be implemented along similar lines as in the NTC scheme described above, by testing only a fraction of all particles for decays. However, since decays only involve one particle in the initial state, the computational cost only grows as $\mathcal{O}(N)$ with the number of particles in the cell. Hence, it is fast enough to iterate over all $\phi$ particles in the cell, compute $P_\textrm{dec}$, generate a random number $u$ and accept a decay if $u<P_\textrm{dec}$. We do this at the beginning of each time step. If a decay happens, the energies of the decay products are given by $E_\rho^*=\frac{m_\phi}{2}$ in the rest frame of the decaying particle and are then boosted into the cosmic frame. We can assume without loss of generality that this boost happens along the $z$-axis, such that the resulting energy of the first $\rho$-particle is given by $E_{\rho,1} = \frac{E_\phi}{m_\phi}E_\rho^*-\frac{p_\phi}{m_\phi}p_\rho^*\cos \theta^*$ where 
$p_\rho^*= \sqrt{(E_\rho^*)^2-m_\rho^2}$ and $\cos \theta^*$ is uniformly distributed in $[-1,1]$ for an isotropic decay. The energy of the first decay product is therefore uniformly distributed between $E_\textrm{min}= \frac{E_\phi}{m_\phi}E_\rho^*-\frac{p_\phi}{m_\phi}p_\rho^*$ and $E_\textrm{max}= \frac{E_\phi}{m_\phi}E_\rho^*+\frac{p_\phi}{m_\phi}p_\rho^*$, while the energy of the second decay product is simply given by $E_{\rho,2}=E_\phi-E_{\rho,1}$.

\paragraph{$2\rho \to 2\rho$ elastic scattering:} The $2\to 2$ scattering cross section is given by 
    \begin{align}
    \nonumber
        (\sigma v)_{2\rho \to 2\rho}&=\frac{1}{4E_1E_2}\frac{1}{2!}
    \int \mathrm{d}\pi_1' \mathrm{d}\pi_2'
\, \left| \mathcal{M}_{2\rho \to 2\rho} \right|^2 
(2\pi)^4
\delta^{(4)}\!\left(p_1 + p_2 - p_1' - p_2'\right)\\
&= \frac{\left| \mathcal{M}_{2\rho \to 2\rho} \right|^2}{4E_1E_2}\frac{1}{16\pi}\sqrt{1-\frac{4m_\rho^2}{s}}=\frac{1}{4E_1 E_2}U_{22}(\sqrt{s}) \; ,
\end{align}
where we defined 
\begin{align}U_{22}(\sqrt{s})&=
\frac{\left| \mathcal{M}_{2\rho \to 2\rho} \right|^2}{16\pi}\sqrt{1-\frac{4m_\rho^2}{s}} \leq  \frac{\left| \mathcal{M}_{2\rho \to 2\rho} \right|^2}{16\pi}\sqrt{1-\frac{4m_\rho^2}{s_\textrm{max}}} = U_{22}^\textrm{max}\end{align}
    with $s_\textrm{max}=4E_\textrm{max}^2$ being the maximal possible CM energy in the simulation volume.
    The flux factor $1/(E_1E_2)$ suppresses scattering of highly-relativistic particles whereas the phase-space factor $(1-4m_\rho^2/s)^{1/2}$ suppresses scattering of non-relativistic particles. Hence, scattering preferentially happens for low-energetic but still relativistic particles. Due to the energy dependence of the upper bound on the cross section, selecting the collision pairs uniformly from the simulation particles creates a large overhead of rejected high-energy pairs. To increase the acceptance rate of collisions, one can preferably select low-energy particles: Instead of the standard NTC algorithm, which requires
    \begin{align}
    N_{\textrm{pairs}}= \binom{N}{2} P_{2\to 2}^\textrm{max}=\frac{1}{2}\sum\limits _{k\neq k'} P^\textrm{max}_{2\to 2}
\end{align}
randomly selected pairs, we follow ref.~\cite{Ihnatenko:2025kew} and select 
\begin{align}
N_{\textrm{pairs}}=\frac{1}{2}\sum\limits _{k\neq k'} P^\textrm{maj}_{k,k'} = \frac{1}{8}\frac{\Delta t}{\Delta x^3}U_{22}^\textrm{max}\left[S_1^2-S_2\right]    
\end{align}
pairs 
with 
\begin{equation}
    P^\textrm{maj}_{k,k'} = \frac{1}{4E_kE_{k'}}\frac{\Delta t}{\Delta x^3}U^\textrm{max}_{22}
\end{equation}
being the energy-dependent majorant on the interaction probability and $S_n=\sum_k (E_k)^{-n}$. Instead of selecting particles randomly with equal weight, these pairs are constructed by selecting the first particle with probability $P=(1/E_1)/S_1$ and the second with probability $P=(1/E_2)/(S_1-1/E_1)$. For each pair, a collision is then accepted with probability $P=U_{22}/U_{22}^\textrm{max}$. For scattering with constant matrix element, this procedure leads to a very high acceptance rate, but it can also be adopted for scattering processes with more general energy dependence.

\paragraph{$2\rho \to 3\rho$ number-changing process:} The cross section is given by 
    \begin{align}
    \nonumber
        (\sigma v)_{2\rho \to 3\rho}&=\frac{1}{4E_1E_2}\frac{1}{3!}
    \int \mathrm{d}\pi_1' \mathrm{d}\pi_2' \mathrm{d}\pi_3'\, \left| \mathcal{M}_{2\rho \to 3\rho} \right|^2 
(2\pi)^4
\delta^{(4)}\!\left(p_1 + p_2 - p_1' - p_2'-p_3'\right)\\
&= \frac{\left| \mathcal{M}_{2\rho \to 3\rho} \right|^2}{4E_1E_2}\frac{1}{3!}\Phi_3(\sqrt{s})\leq \frac{\left| \mathcal{M}_{2\rho \to 3\rho} \right|^2}{4E_1E_2}\frac{1}{3!}\Phi_3\left(\sqrt{2m_\rho^2+ 2(E_1E_2 + p_1p_2)}\right)
\end{align}
with the 3-particle phase space $\Phi_3(\sqrt{s})$. For the inequality, we use the monotonicity of $\Phi_3(\sqrt{s})$ to substitute the largest possible CM energy for given energies $E_1$ and $E_2$ of the $\rho$ particles in the cosmic frame. In the relativistic limit, the 3-particle phase space is given by $\Phi_3(s)=s/(256\pi^3) \leq 4 E_1 E_2 / (256\pi^3)$, so that
\begin{align}
(\sigma v)_{2\rho \to 3\rho}^\textrm{max}=\frac{1}{3!}\frac{\left| \mathcal{M}_{2\rho \to 3\rho} \right|^2}{256\pi^3} \; .
\end{align}

We emphasise that our assumption of a constant matrix element implicitly corresponds to an effective interaction of the form $\tfrac{1}{\Lambda}\rho^5$ where $\Lambda$ has dimension of energy. The squared matrix element is then proportional to $\Lambda^{-2}$, which has the unit of a cross section, such that no additional factors of energy are needed. Compared to the case of $2\to2$ scattering (where the squared matrix element is dimensionless), this leads to an enhancement of the scattering probability for highly-energetic particles, as expected for a non-renormalisable interaction.

\paragraph{$3\rho \to 2\rho$ number-changing scattering:} The scattering probability given in eq.~\eqref{eq: scattering prob 3 to 2} depends on the factor $I_{32}/(E_1E_2E_3)$. For a constant matrix element, $I_{32}$ reduces to 
\begin{align}
    I_{32}= \frac{\left| \mathcal{M}_{3\rho \to 2\rho} \right|^2}{16\pi} \sqrt{1-\frac{4m_\rho^2}{s}} \; .
\end{align}
The CM energy satisfies $9 m_\rho^2 \leq s < \infty$, such that $\sqrt{1 - 4 m_\rho^2 /s} \in (0.74, 1)$ and the values of $I_{32}$ have almost no dependence on the scattering kinematics. Most importantly, we have
\begin{align}
I_{32} \leq \frac{\left| \mathcal{M}_{3\rho \to 2\rho} \right|^2}{16\pi} = I_{32}^\textrm{max} \; .
\end{align}
The interaction probability is therefore mostly determined by the energy-dependent factor $1/(E_1 E_2 E_3)$, leading to enhanced scattering probability for low-energy particles. We adopt a similar approach as for the case of $2 \to 2$ scattering and select 
\begin{align}
N_{3-\textrm{plets}}=\frac{1}{6}\sum\limits _{k\neq k'\neq k''} P^\textrm{maj}_{k,k',k''} = \frac{1}{48}\frac{\Delta t}{(\Delta x^3)^2}I_{32}^\textrm{max}\left[S_1^3-3S_1S_2+2S_3\right]    
\end{align}
pairs with $P^\textrm{maj}_{k,k',k''} = \frac{1}{8E_kE_{k'}E_{k''}}\frac{\Delta t}{(\Delta x^3)^2}I^\textrm{max}_{32}$ and $S_n$ as defined above. We select the first particle of the triplet with probability $P=(1/E_1)/S_1$, the second with probability $P=(1/E_2)/(S_1-1/E_1)$ and the third with probability $P=(1/E_3)/(S_1-1/E_1-1/E_2)$. A triplet collision is then accepted with probability $P=I_{32}/I_{32}^\textrm{max}$.

\paragraph{Sampling of final states:} In case a collision is accepted, a random final state must be sampled. The differential collision probability for $n\to m$ scattering is $\mathrm{d}P_{n\to m}\propto \mathrm{d}\Phi_m|\mathcal{M}_{n\to m}|$, where $\mathrm{d}\Phi_m$ is the differential $m$-body phase space. Since we assume for simplicity constant matrix elements for all processes, the final state of a collision is sampled by first going to the CM frame of the incoming particles, computing the CM-energy, then uniformly sampling one $m$-particle final state from the $m$-particle phase space for the given CM-energy and then boosting this final state back into the cosmic frame. For a $2$-particle final state, the uniform phase-space sampling is done by calculating the final state energies in the CM-frame and then sampling the polar and azimuthal scattering angles $\cos(\theta)$ and $\varphi$ uniformly from $[-1,1]$ and $[0,2\pi)$. For the 3-particle final state, we use the RAMBO algorithm as implemented in the HAZMA package~\cite{Kleiss:1985gy,Coogan:2019qpu,Coogan:2022cdd}. 

\paragraph{Multiple interactions: }
During each time step, the different types of interaction can happen simultaneously. We therefore compute the number of tested pairs for all the different interaction types and sum them to $N_\textrm{total}$. Then, we draw a number from a Poisson distribution with mean $N_\textrm{total}$ and iterate up to this number. At each iteration we choose one of the scattering processes with the weight given by its contribution to $N_\textrm{total}$ and then perform the stochastic collision test.

\subsection{Evaluation of results}

From our DSMC simulations, we want to extract various quantities that can have an observable impact on the evolution of the universe. The simplest such quantity is the particle yield $Y = n / s$, where $s$ is the entropy density of the Standard Model thermal bath with the entropy effective degrees of freedom assumed to be constant at $g_{*,S}=3.91$. For cosmologically stable particles, the yield is expected to approach a constant value $Y_0$ for $T \to 0$, which can be compared to the measured abundance of DM in the present universe. The yield is obtained by simply counting the number of particles in the simulation volume and dividing by the entropy density: $Y = N / (V(t) \cdot s(t))$.

Another quantity of interest is the temperature $T_x$ of the particle species $x$, which gives a first indication of whether the particles will cluster like cold or warm DM during structure formation. Determining the temperature requires the calculation of an expectation value, see eq.~\eqref{eq: effective temperature definition}. Here we once again benefit from the Monte Carlo approach, because we can replace the integral $\int \frac{\mathrm{d}^3p}{(2\pi)^3} f(p)$ simply by an ensemble average $(1/V)\sum_i$. The temperature is therefore calculated as
\begin{equation}
    T_x= \bigg\langle \frac{p^2}{3 E} \bigg\rangle_x = \frac{1}{N}\sum_i \frac{p_i^2}{3 E_i} \; . 
\end{equation}

\subsection{Validation tests}
\label{sec:validation}

In order to confirm the implementation of the algorithm presented above, we have performed multiple tests, which we briefly present in the following.

\begin{figure}[t]
\centering
\includegraphics[width=\textwidth]{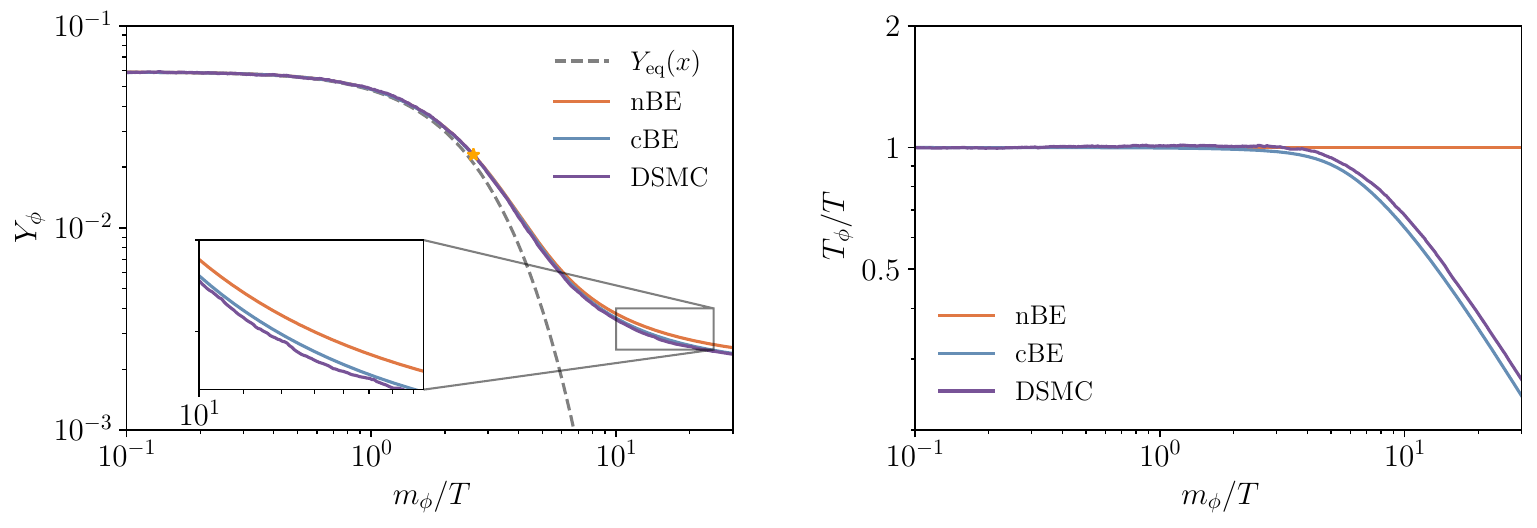}
\caption{Left: Evolution of the $\phi$ yield during freeze-out. Shown are results of the nBE, cBE and DSMC approaches. The star indicates where the nBE solution deviates by 10\% from the equilibrium yield indicated by the dashed line. Right: Evolution of the $\phi$ temperature compared to the SM bath-temperature for cBE and DSMC approaches. \label{fig: freeze out check}}
\end{figure}

\paragraph{$\psi \psi \leftrightarrow \phi \phi$ annihilation:} We perform a simulation for the $\psi \psi \leftrightarrow \phi \phi$ scattering with $|\mathcal{M}_{\psi \psi \leftrightarrow \phi \phi}|^2= 3.4\times 10^{-11}$ and show the evolution of the $\phi$ yield and temperature in figure~\ref{fig: freeze out check}. We compare our simulation to the standard freeze-out computation with the nBE approach (see section~\ref{sec:set-up}). However, since we do not include elastic scattering, $\psi \phi \to \psi \phi$, the usual assumption that $\phi$ is in kinetic equilibrium with the thermal bath throughout the freeze-out procedure is not valid and hence we expect the standard nBE approach for freeze-out to deviate from our simulation. We therefore also compare to the cBE-approach for freeze-out, which is suitable for the case of early kinetic decoupling because it allows for a separate temperature of the $\phi$ particles.

The left panel of figure~\ref{fig: freeze out check} shows the well-known freeze-out behaviour as the $\phi$ particles decouple from chemical equilibrium. While there is a clear difference between the DSMC result and the nBE approach, the DSMC and the cBE approach agree very well. The right panel of figure~\ref{fig: freeze out check} shows how the temperature $T_\phi$ starts to deviate from the temperature $T$ of the thermal bath. This deviation arises because the production of highly energetic $\phi$ particles from the thermal bath is suppressed, such that $T_\phi$ decreases more quickly. The small difference between the temperature determined from the DSMC and from the cBE approach arises because the $\phi$ particles are not in kinetic equilibrium with themselves after the decoupling, i.e.\ the assumption of a thermal phase-space distribution made in the cBE approach is not perfectly valid in the absence of $\phi \phi \to \phi \phi$ scattering processes. 

For the matrix element chosen in figure \ref{fig: freeze out check}, the $\phi$ particle yield is suppressed by less than two orders of magnitude before freeze-out occurs. Larger $\psi \psi \leftrightarrow \phi \phi$ annihilation rates would correspond to a later freeze-out and hence a stronger suppression. If the yield drops over many orders of magnitude, it becomes difficult to perform the DSMC, because one would need to start with a huge number of particles during the initial stages in order to end up with a statistically significant number of $\phi$ particles after decoupling. Simulating such a large number of particles with larger cross sections becomes computationally infeasible. To circumvent this difficulty for our simulations in section~\ref{sec: results}, we make use of the fact that in the initial stages of freeze-out, the $\phi$ momentum distribution still pretty accurately tracks the thermal equilibrium distribution and $T_\phi \approx T$. Instead of performing a simulation that captures the full $\psi \psi \leftrightarrow \phi \phi$ decoupling with the initial equilibrium stage, we will start the simulations presented in section \ref{sec: results} when the nBE solution deviates from the equilibrium yield $Y_\textrm{eq}$ by 10\%. This point is marked by an orange star in figure~\ref{fig: freeze out check} for the given freeze-out process. Until this point we simply sample the $\phi$ particles from a Maxwell-Boltzmann distribution with SM-bath temperature.

\paragraph{$\phi \to \rho \rho$ decay:} For a decaying particle $\phi$, which follows a Maxwell-Boltzmann distribution, $f_\phi(p)\sim \exp(-E/T_\phi)$, the Boltzmann equation for decay can be solved analytically \cite{Hall:2009bx}. The resulting yield of $\rho$ particles is given by 
\begin{align}
    Y(T_\phi) = \frac{135}{\pi^5}\sqrt{\frac{5}{2}}\frac{M_\textrm{pl}\Gamma_\textrm{dec}}{m_\phi^2g_*^S\sqrt{g_*}}\int_0^{m_\phi/T_\phi} \mathrm{d}x\, x^3 K_1(x)
\end{align}
with $K_1$ being the first modified Bessel function of the second kind. Furthermore, the distribution function of $\rho$ momenta after the completion of in-equilibrium freeze-in for $m_\rho \ll m_\phi$ is given by 
\begin{align}
    f_\rho(p) = 12  \sqrt\frac{10T_\phi\Delta}{ g_* \pi p} \frac{M_\textrm{pl}\Gamma_\textrm{dec}}{m_\phi^2\Delta^3}\exp\left(-\frac{p}{T_\phi\Delta }\right)
\end{align}
with $\Delta = 1-m_\rho^2/m_\phi^2$. 

\begin{figure}[t]
\centering
\includegraphics[width=\textwidth]{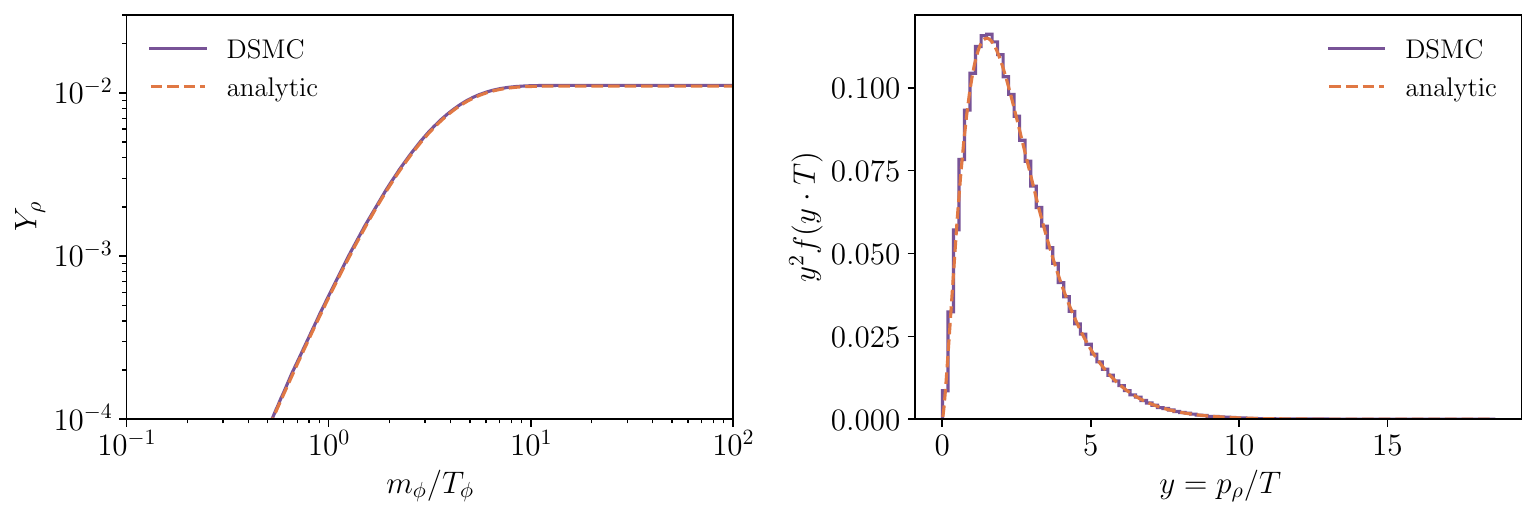}
\caption{Left: Evolution of the $\rho$ yield from in-equilibrium decay of $\phi$ particles. Right: Distribution of $\rho$ momenta normalised to temperature. \label{fig: decay check}}
\end{figure}

In figure~\ref{fig: decay check}, we show the evolution of the $\rho$ yield and the late-time momentum distribution for the DSMC together with the analytical solution for a benchmark where $\Gamma_\textrm{dec} = 10^{-14}\, \textrm{GeV},\, m_\phi = 10^3 \, \textrm{GeV},\, m_\rho = 10\, \textrm{keV}$. The DSMC agrees very well with the analytic solution. 

\paragraph{$2\leftrightarrow3$ number-changing scattering:}
For $2\leftrightarrow 3$ scattering, we can check that in a static universe, number-changing processes lead to thermal equilibrium. Since we assume a constant matrix element, all momenta above the kinematic threshold are involved in these processes, and hence we expect that both kinetic and chemical equilibrium are eventually reached. Given the initial energies of the particles in the simulations, we can then determine the number density and temperature in the equilibrium state using energy conservation in a non-expanding universe: Neglecting effects from quantum statistics, kinetic equilibrium leads to a phase-space distribution of the shape $f(p) = \exp(-(E-\mu)/T)$ with the chemical potential $\mu$. In chemical equilibrium, we furthermore have $\mu=0$, such that the number and energy densities are solely a function of the equilibrium temperature:
\begin{align}
    \label{eq: equilibrium number density}
    n(T) &= \frac{m^2T}{2\pi^2}K_2\left(\frac{m}{T}\right) \; ,\\
    \label{eq: equilibrium energy density}
    \rho(T) &=\frac{m^2T^2}{2\pi^2}\left(3K_2\left(\frac{m}{T}\right) + \frac{m}{T} K_1\left(\frac{m}{T}\right)\right) \; .
\end{align}
From the known initial energy density, we can thus compute the equilibrium temperature using eq.~\eqref{eq: equilibrium energy density} and then the equilibrium number density using eq.~\eqref{eq: equilibrium number density}.  

As initial distribution, we choose a Maxwell-Boltzmann distribution with temperature $T_\textrm{initial}= 5$.\footnote{Since there is no reference scale in the absence of expansion, we use dimensionless parameters in this paragraph} We start with $N=8 \cdot 10^4$ particles in a volume $V=800$. The mass of the $\rho$ particles is chosen to be $m_\rho = 1$ and the matrix element is $\left| \mathcal{M}_{2\rho \to 3\rho} \right|^2 = 1$. For these initial conditions, the initial value of $n$ is larger than its value in chemical equilibrium. In other words, we expect $3 \to 2$ processes to dominate over $2 \to 3$ processes, leading to a decrease in particle number and an increase in temperature as the system approaches chemical equilibrium.\footnote{For such a highly occupied state, quantum statistics are generally not negligible. For the purpose of validating our implementation it is however sufficient to proceed with classical particles.}

\begin{figure}[tbp]
\centering
\includegraphics[width=\textwidth]{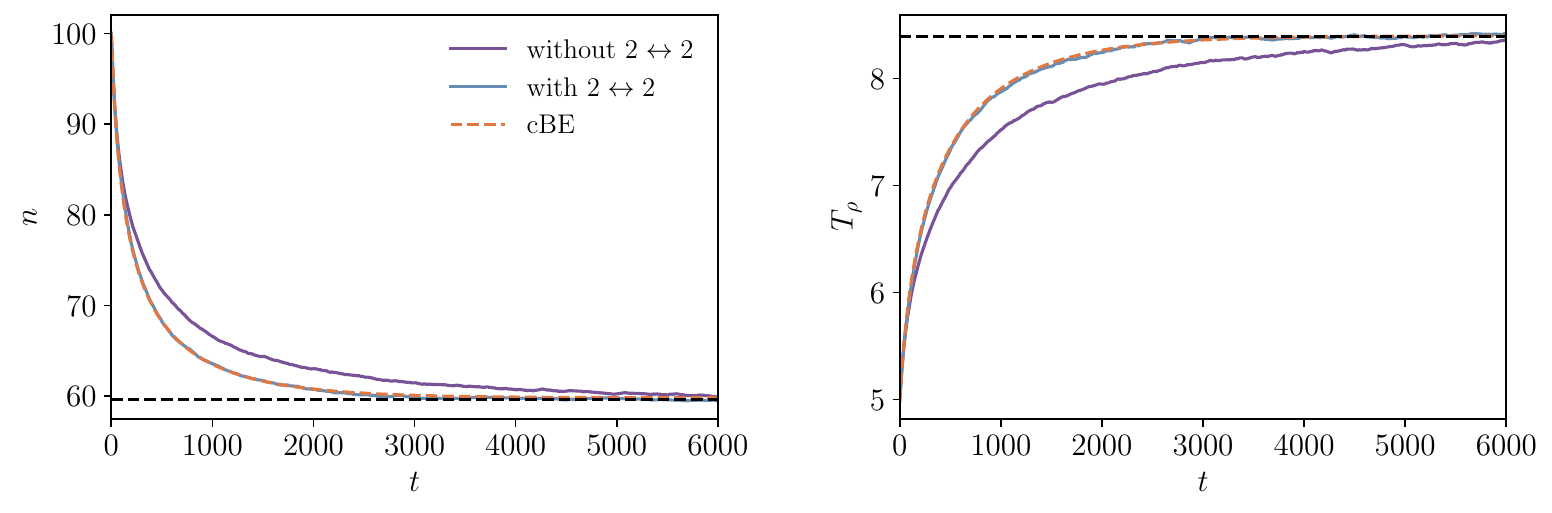}
\caption{Evolution of number density (left) and temperature (right) towards equilibrium. The results with efficient $2\to 2$ scattering are shown in blue, without $2\to 2$ scattering in purple. The black dashed line shows the equilibrium quantities, the dashed orange line the cBE-solution.  \label{fig: 3 to 2 check}}
\end{figure}

\begin{figure}[tbp]
\centering
\includegraphics[width=\textwidth]{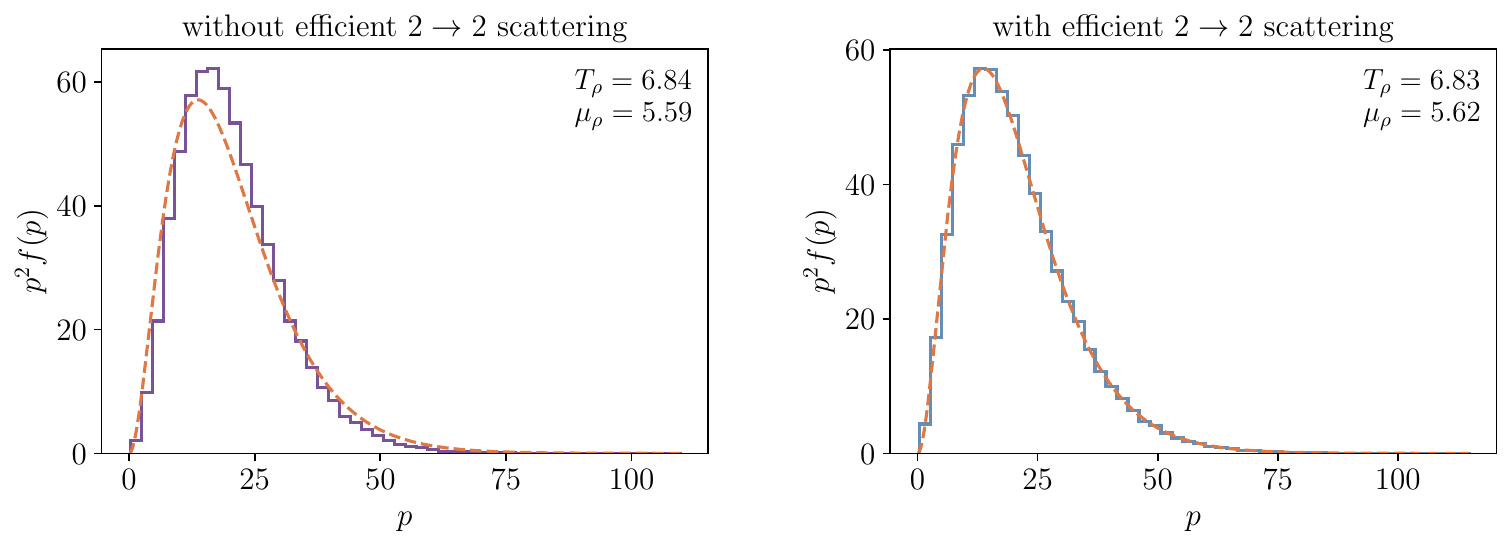}
\caption{Momentum histogram at intermediate temperatures with and without efficient $2\to 2$ scattering together with the kinetic equilibrium distribution (yellow) for the corresponding temperature and chemical potential of the ensemble.\label{fig: histo 3 to 2 check}}
\end{figure}

In figure \ref{fig: 3 to 2 check} we show the evolution of number density and effective temperature under $2\leftrightarrow3$ processes. The purple line represents the scenario where only $2\leftrightarrow 3$ collisions are present whereas the blue line includes efficient $2\leftrightarrow 2$ collisions to keep the distribution in kinetic equilibrium throughout the simulation. 
We see that in both cases, number density and effective temperature approach their equilibrium values. The dashed orange line in figure \ref{fig: 3 to 2 check} shows the solution to the cBE-approach described in section \ref{ch: phase space}, which matches the DSMC simulation in case of efficient $2\to2$ collisions. This result agrees with expectation, since the cBE approach assumes kinetic equilibrium, which is ensured by the elastic scattering processes.

When only $2\leftrightarrow 3$ collisions are included, the DSMC solution deviates from the cBE solution. This is because the $2\leftrightarrow 3$ collisions are not fast enough to maintain kinetic equilibrium for changing $n$. This result is visualised in figure \ref{fig: histo 3 to 2 check}, where we show the distribution of momenta at an intermediate time together with the kinetic equilibrium shape given the effective temperature and chemical potential of the particle ensemble. For the case of only $2\leftrightarrow 3$ collisions, we see departures from the Maxwell-Boltzmann shape at intermediate times, whereas with efficient $2 \to 2$ collisions, the distribution of ensemble momenta closely tracks the Maxwell-Boltzmann shape at all times. 

\section{Results}
\label{sec: results}

Having validated our implementation, we now present the results from our simulations of the toy model described in section~\ref{subsec toy model}. Our benchmark parameters are $m_\phi = 1 \, \textrm{TeV},\, m_\rho = 10\, \textrm{keV}, \Gamma_\textrm{dec} = 10^{-17} \, \textrm{GeV},$ $ \left| \mathcal{M}_{\psi\psi \to \phi \phi} \right|^2 = 2.3
\times 10^{-8}$. These parameters are chosen such that the different processes happen at different periods in the early universe: The freeze-out of $\phi$ happens when the particles become Boltzmann-suppressed at $T \sim 100 \, \mathrm{GeV}$, while out-of-equilibrium decays happen when $\Gamma_\text{dec} \gtrsim H$ at $T \lesssim 10 \, \mathrm{GeV}$. The $\rho$ particles are still highly relativistic at this point and the precise value of their mass plays no role in the following discussion. The value $m_\rho = 10 \, \mathrm{keV}$ is chosen such that the observed DM relic abundance $\Omega h^2 = 0.12$ is reproduced in the absence of number-changing processes. This value also lies in the mass range where warm DM constraints become relevant. 
We start the simulation at a bath temperature of $T_0=10m_\phi$. The simulation volume $V_i$ at the initial temperature is $V_i = 10^{-4}\, \textrm{GeV}^{-3}$. This corresponds to a volume of roughly $60\,\textrm{L}$ today and $\mathcal{O}(7000)$ $\rho$ particles after the freeze-in is finished in the case of no number-changing processes. We chose the time step size as $\Delta t = 0.01\times\min (H^{-1}, \Gamma_\textrm{max}^{-1})$, with $\Gamma_\textrm{max}$ denoting the maximal interaction rate of all possible interactions.  

\begin{figure}[tbp]
\centering
\includegraphics[width=\textwidth]{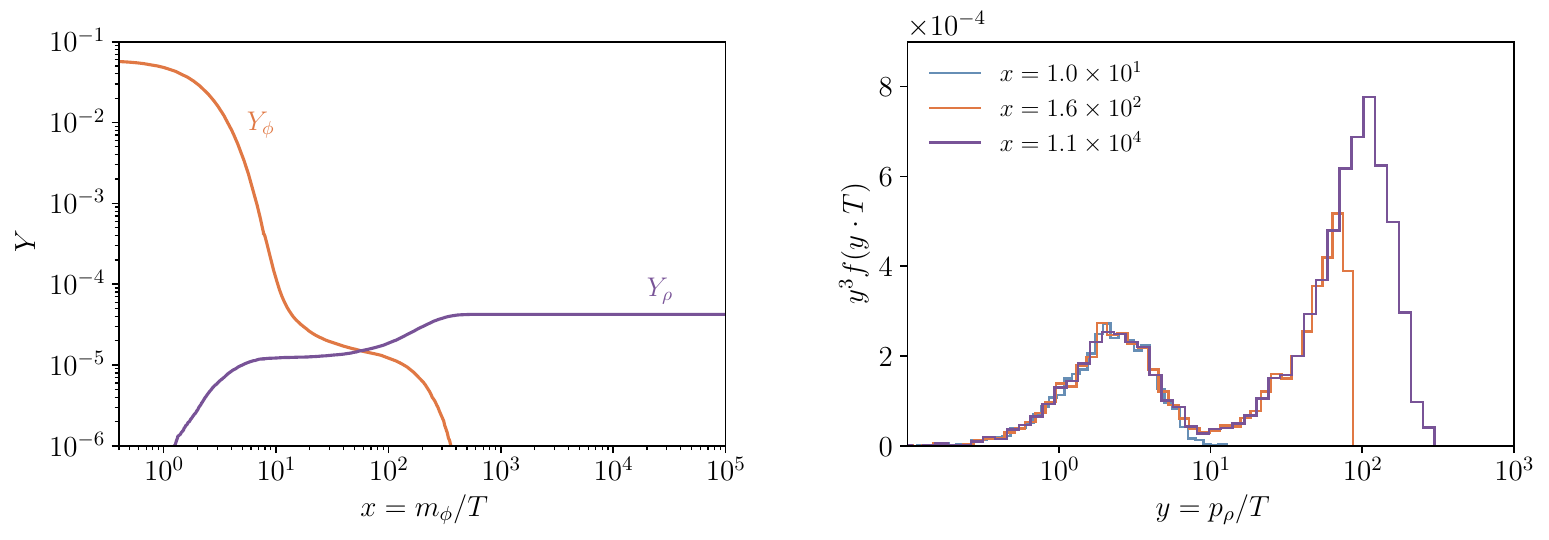}

\caption{Left: Evolution of the $\rho$ yield from decays of $\phi$ particles. Right: Distribution of $\rho$-momenta at different times normalised to temperature. \label{fig: baseline freeze in} }
\end{figure}

Before including number-changing processes, we first present the pure freeze-in simulation results as a baseline. We show the evolution of the yields for $\phi$ and $\rho$ as well as the evolution of the phase-space distribution of $\rho$ in figure~\ref{fig: baseline freeze in}. We plot the phase-space distribution in terms of $p_\rho / T$, which is invariant under the expansion of the universe as long as $g_\star$ for the SM thermal bath is approximately constant such that $T \propto a^{-1}$. Around $m_\phi/T \sim 3-5$, in-equilibrium decays of the $\phi$ particles start to populate the $\rho$ sector. This creates the low-momentum peak at $p_\rho/T\sim 2.5$ \cite{Heeck:2017xbu} in the right part of figure~\ref{fig: baseline freeze in}. When the $\phi$ particles freeze out and decay, they create the high-momentum peak at $p_\rho/T\sim m_\phi/(2T_\textrm{dec})$, where $T_\textrm{dec}$ is defined by  $H(T_\text{dec}) = \Gamma_\textrm{dec}$.

\subsection{$2 \rho \to 2 \rho$ scattering}

\begin{figure}[tbp]
\centering
\includegraphics[width=\linewidth]{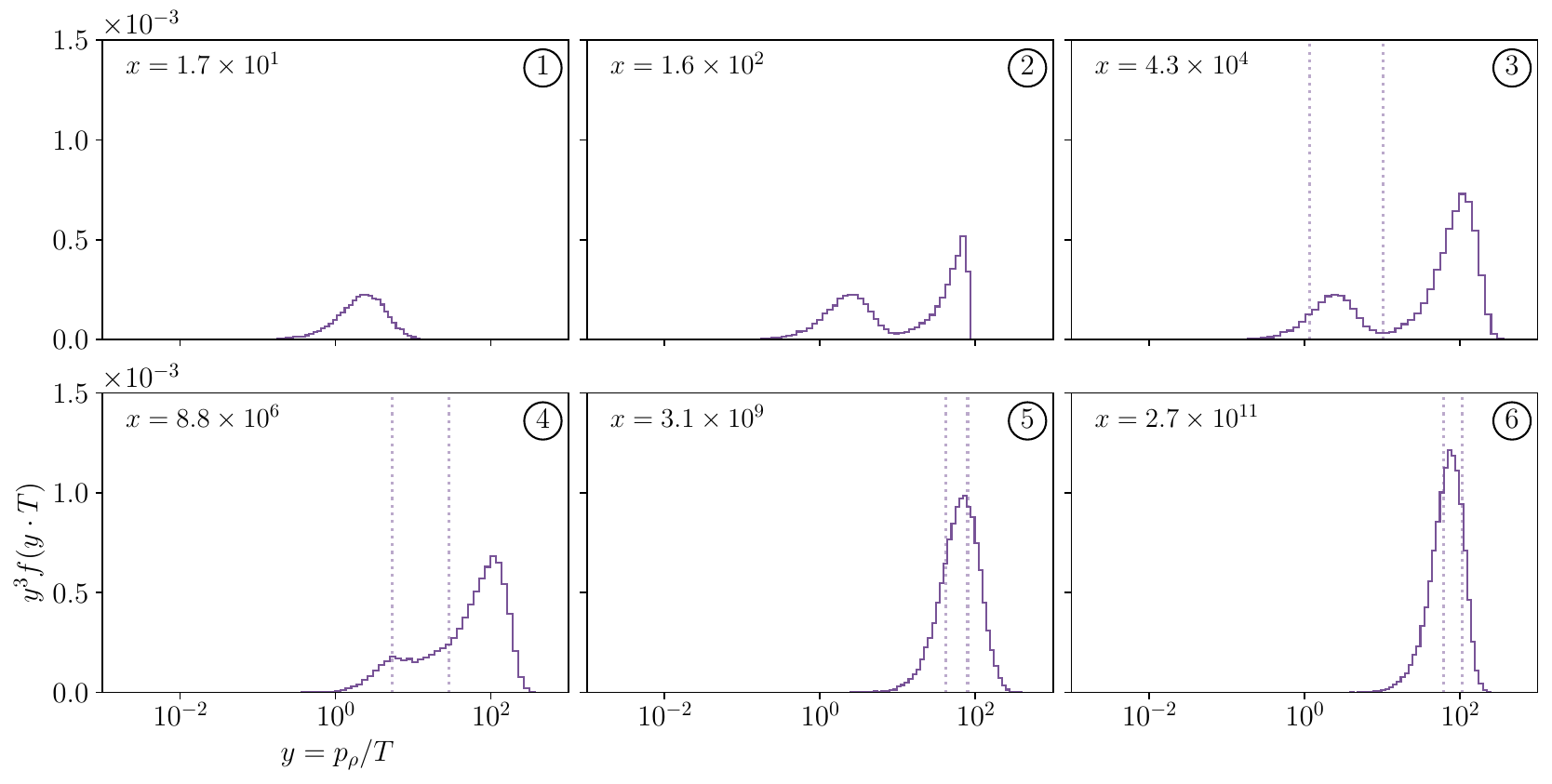}
\caption{Evolution of the phase-space distribution under $2\to 2$ scattering. The vertical dashed lines indicate the average momenta of the low and high energy scattering partners in the collision.}    \label{fig:2to2_histogram_evolution}

\vspace{5mm}

\includegraphics[width=\textwidth]{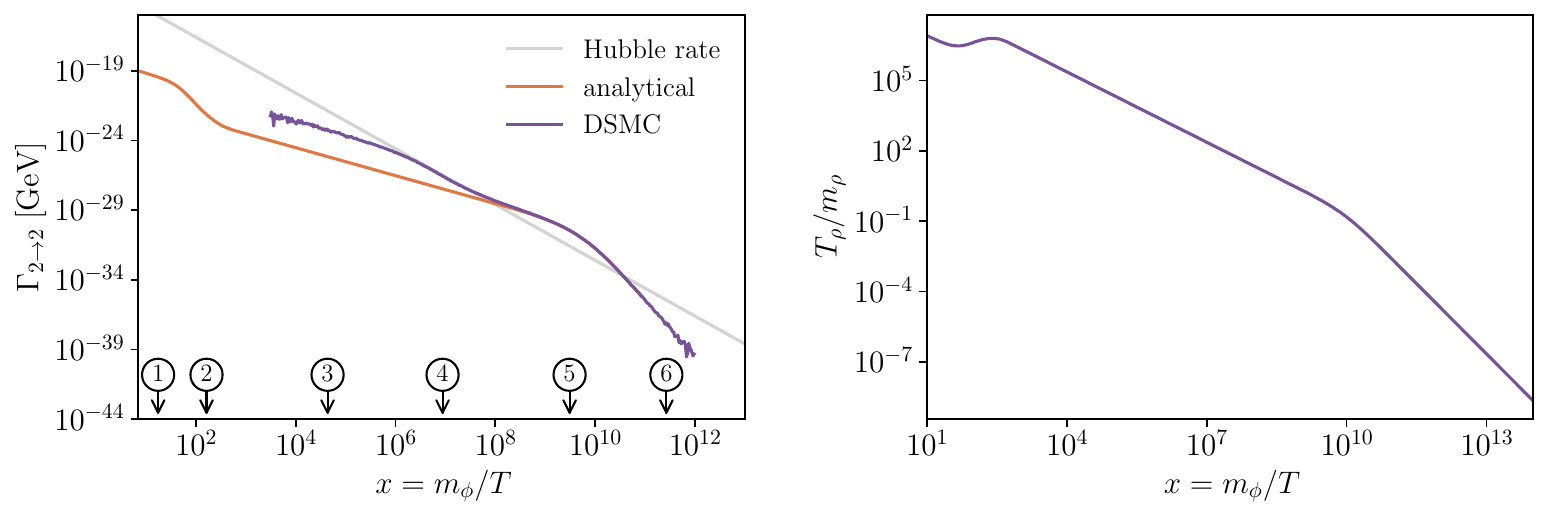}
\caption{Left: Evolution of the $2\to 2$ scattering rate in the simulation together with the analytical approximation assuming kinetic equilibrium, and the Hubble rate. Right: Evolution of the effective $\rho$ particle temperature $T_\rho$ in the simulation. \label{fig: temperature and rate 2 to 2}}
\end{figure}

We now include the $2\rho \to 2\rho$ scattering with a matrix element $|\mathcal{M}_{2\rho\to 2\rho}|^2 = 2\times 10^{-14}$, corresponding to a self-scattering cross section of $\sigma/m_\rho = |\mathcal{M}_{2\rho\to 2\rho}|^2/(128\pi m_\rho^3)\sim10^{-5}\, \textrm{cm}^2/\textrm{g}$ in the non-relativistic limit, and an initial volume $V= 10V_i$.
Including $2\rho \to 2\rho$ scattering leads to reshuffling of momenta in the $\rho$ phase-space distribution but not to changes in the number density. Hence the yield evolution remains the same as in the left side of figure \ref{fig: baseline freeze in}.
We show the simulation results for the evolution of the phase-space distribution including $2\rho \to 2\rho$ scattering in figure \ref{fig:2to2_histogram_evolution}. In figure \ref{fig: temperature and rate 2 to 2}, we show on the left the evolution of the $2\to 2$ scattering rate per particle, 
\begin{align}
    \Gamma_{2\to2} = 2\times \frac{\text{number of }2\to 2\text{ collisions}}{N_\rho \Delta t} \; ,
\end{align}
along with the analytical expectation in case of kinetic equilibrium. 
The analytical prediction is given by $\Gamma^\text{eq}_{2\to 2} = n_\rho \langle \sigma v \rangle^\text{eq}_{2\to 2} (T_\rho)$ with the thermally averaged cross section for general $n\rho\to m\rho$ scattering
\begin{align}
    \label{eq: thermally averaged cross section definition}
    \langle \sigma v^{n-1} \rangle^\text{eq}_{n\rho\to m\rho}
    &= \frac{1}{n_\textrm{eq}(T_\rho)^n}\frac{1}{m!}\int \prod_{\substack{i\in n\\ j\in m}} \mathrm{d}\pi_i \mathrm{d}\pi_j'|\mathcal{M}_{n\rho\to m\rho}|^2f_i^\textrm{eq}(T_\rho)(2\pi)^4 \delta^{(4)}\Bigl(\sum_{i\in n}p_i-\sum_{j \in m} p_j'\Bigr),
\end{align}
where $f^\textrm{eq}(p, T)=\exp(-E/T)$ and $n_\textrm{eq}(T)=\int \textrm{d}^3 p/(2\pi)^3 f^\textrm{eq}(p,T)$. We use the effective $\rho$ particle temperature obtained from our simulation (c.f.\ eq.~\eqref{eq: effective temperature definition}, shown on the right side of figure \ref{fig: temperature and rate 2 to 2}) as input for the analytical scattering rate.
 
The evolution of the phase-space distribution can be understood by looking at the actual collision rate obtained from the simulation compared with the analytical prediction assuming kinetic equilibrium. The following picture emerges:
\begin{itemize}
\item[\textcircled{1}] Up to $x=m_\phi/T \sim 40$, the $\rho$-population consists only of the low energy peak created through the in-equilibrium decay. Since the particles are all highly relativistic and the cross section scales as $\sigma(\sqrt{s}) \propto s^{-1}$, $2\to 2$ scattering is suppressed. In fact, the scattering rate is so small that it cannot be reliably estimated from our simulations and is therefore not shown in figure~\ref{fig: temperature and rate 2 to 2}.

\item[\textcircled{2}] Between $m_\phi/T = 40$ and $600$, the out-of-equilibrium decays of the $\phi$ particles happen, producing the higher-momentum peak in the distribution. This leads to the visible increase in the effective temperature $T_\rho$ seen in the right panel of figure \ref{fig: temperature and rate 2 to 2}. The increase of the effective temperature leads to a decrease in the would-be equilibrium scattering rate as higher energy particles are less likely to scatter. All momenta of the particles are still too large to lead to effective scattering.

\item[\textcircled{3}] As the universe expands, the momenta redshift and the lower momentum peak reaches the momentum range where scattering is no longer negligible. The fact that the scattering rate in the simulation is larger than the equilibrium estimate reflects the highly non-thermal shape of the true distribution. Since the true distribution is broader than the equilibrium one, and since $\sigma$ is a convex function of $\sqrt{s}$, it follows that $\langle \sigma v \rangle > \langle \sigma v \rangle^\text{eq}$.
Below $m_\phi/T \sim 10^6$, the scattering rate is still smaller than the Hubble rate and therefore does not efficiently alter the phase-space distribution. However, the interaction rate for relativistic particles decreases as $\Gamma \propto T$, so slower than the Hubble rate. 

\item[\textcircled{4}]  As soon as the scattering rate becomes comparable to the Hubble rate, kinetic equilibration at the lower end of the momentum spectrum becomes efficient. The scattering processes slowly alter the momentum distribution towards the equilibrium one and thereby lower the scattering rate towards the equilibrium value. During this approach to equilibrium, the scattering rate approximately tracks the Hubble rate. The fact that the two rates coincide almost exactly for an extended period of time is to some degree a coincidence for the chosen separation and heights of the momentum peaks. For different parameter values, the scattering rate does not exactly follow the Hubble rate, but the qualitative behaviour of the two rates scaling in a similar way remains. 

\item[\textcircled{5}] As the momenta redshift further, also the high momentum peak becomes relevant for scattering. Eventually, also the equilibrium scattering rate becomes larger than the Hubble rate, at which point the distribution reaches full kinetic equilibrium, such that the actual scattering rate tracks the analytical estimate perfectly.

\item[\textcircled{6}] At around $m_\phi/T = 10^{10}$, the effective temperature drops below the mass. As a result, the momentum distribution changes from the relativistic form, $f(p) \propto p^2 e^{-p/T}$, to the non-relativistic one, $f(p) \propto p^2 e^{-p^2 / (2 m_\rho T)}$. The scattering cross section approaches a constant in the non-relativistic limit but number density and the relative velocities continue to decrease. Soon after the particles have become non-relativistic, the scattering therefore becomes inefficient and the interaction freezes out.
\end{itemize}   
For the chosen parameter values, the DM particles become non-relativistic only for very small temperatures of the SM thermal bath, $T \lesssim 0.1 \, \mathrm{keV}$. This value results from the small assumed DM mass $m_\rho = 10 \, \mathrm{keV}$ and the large momenta injected by out-of-equilibrium decays ($p_\rho / T \sim 100$). As a result, the model considered here clearly does not satisfy the typical warm DM bounds and would not give successful structure formation~\cite{Egana-Ugrinovic:2021gnu}. This finding motivates a detailed investigation of the effect of number-changing processes which lead to a cooling of the dark sector~\cite{Cervantes:2026cfq}.

\subsection{$2\rho \leftrightarrow 3\rho$ number-changing scattering}
\begin{figure}[htbp]
\centering

\includegraphics[width=\textwidth]{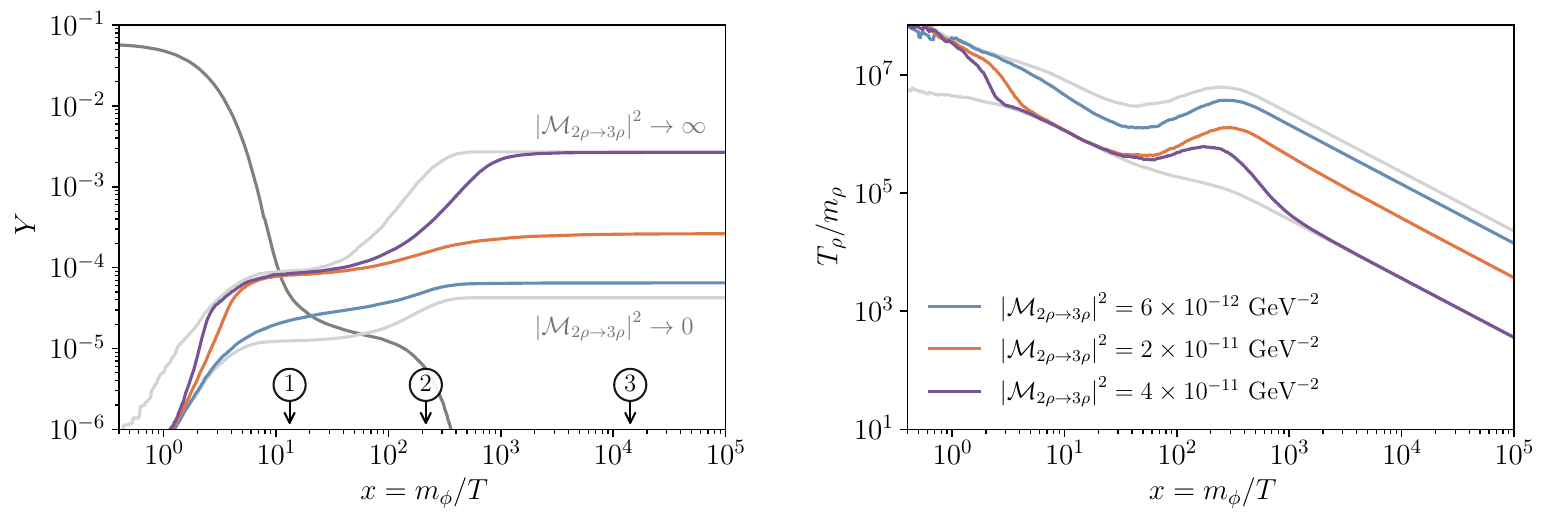}
\caption{ Evolution yield (left) and effective temperature (right) for different $2\leftrightarrow 3$ matrix elements.\label{fig:  yield and temperature evolution numberchanging scattering}}

\vspace{5mm}

\includegraphics[width=\textwidth]{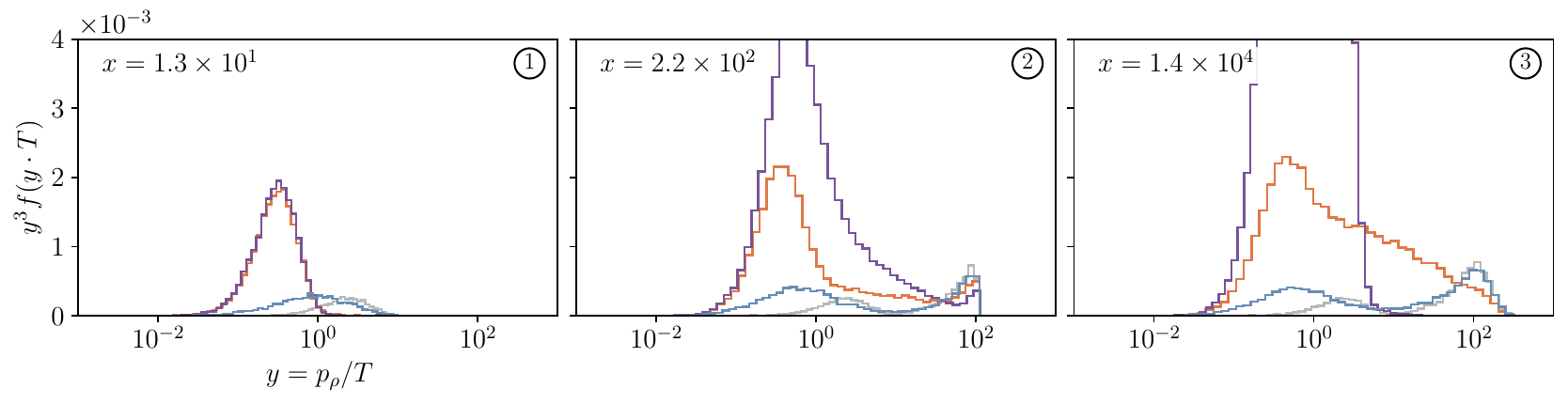}
\caption{Evolution of the momentum distribution under $2\leftrightarrow 3 $ scattering for the matrix elements of the simulations shown in figure~\ref{fig:  yield and temperature evolution numberchanging scattering}. The grey contour shows the momentum distribution in absence of self-scattering.  \label{fig:  evolution momentum distribution with  2 and 3}}
\end{figure}

Let us now turn to the case with number-changing processes. In figure~\ref{fig:  yield and temperature evolution numberchanging scattering}, we show the evolution of the $\rho$ yield and effective $\rho$ temperature of the simulated particle ensemble for various $2\rho \leftrightarrow 3\rho$ matrix elements. The grey-shaded area corresponds to all possible values between no number-changing interactions at all (as considered above) and full chemical equilibrium. The latter case can be readily calculated for relativistic particles, since the energy density redshifts as $\rho \propto a^{-4}$, so we can calculate the equilibrium quantities using energy conservation as discussed in section~\ref{sec:validation}. Crucially, we find that the set-up that we consider corresponds to an under-occupied dark sector at the end of the freeze-in stage, such that $2 \to 3$ processes dominate over $3 \to 2$ processes, leading to an increase in the number density of $\rho$ particles.

As expected, when increasing the matrix elements for number-changing processes, the $\rho$ particle yield is driven more and more towards its chemical equilibrium value. The increase in number density leads to a cooling of the dark sector (see the right panel of figure~\ref{fig:  yield and temperature evolution numberchanging scattering}) as kinetic energy is converted into the rest mass of additional particles. We further observe that thermalisation during the period of in-equilibrium decays ($m_\phi / T \sim 10$) requires smaller matrix elements than thermalisation during the period of out-of-equilibrium decays ($m_\phi / T \sim 100\text{--}1000$).

In figure~\ref{fig:  evolution momentum distribution with  2 and 3} we show the phase-space distribution at several points in the evolution for the different matrix elements. The shown stages correspond to the following behaviour:
\begin{itemize}
    \item[\textcircled{1}] For large matrix elements, the in-equilibrium decay abundance thermalises directly, whereas the smallest matrix element does not lead to complete thermalisation. Compared to the phase-space distribution without number-changing processes (shown in grey), the $2\to 3$ processes increase the total number of particles and distribute the particle momenta towards smaller values.
    \item[\textcircled{2}] During the out-of-equilibrium decays, the second peak is produced. For the larger matrix elements, the number-changing processes drive the distribution towards equilibrium. For the smallest matrix element, the interaction is no longer efficient and the high-energy peak does not thermalise at all. 
    \item[\textcircled{3}] The out-of-equilibrium decay has finished and the $\rho$ particle yield has become constant. For the largest matrix element, complete thermal equilibrium was achieved. The intermediate matrix element was not large enough to achieve complete thermal equilibration for the second peak, and for the smallest matrix element, the second peak is unchanged from the decays. 
\end{itemize}

\begin{figure}[tbp]
\centering
\includegraphics[width=\textwidth]{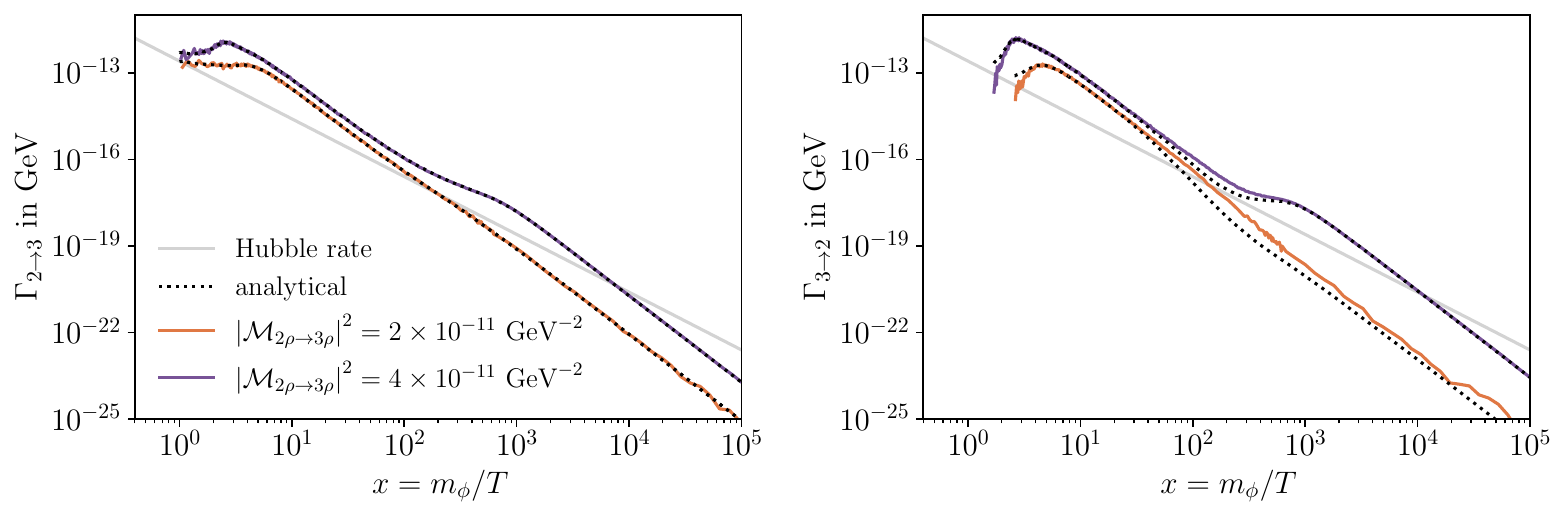}
\caption{$2\to 3$ (left) and $3\to 2$ (right) scattering rates for the two largest matrix elements of the simulations shown in figure \ref{fig:  yield and temperature evolution numberchanging scattering}. We do not include the rate for the lowest matrix element here as scattering processes happen too rarely to calculate the corresponding rate. The dotted line shows the analytic expectations in case of thermal equilibrium.\label{fig:  2 to 3 scattering rates} }
\end{figure}

To understand these findings, we can look at the per-particle interaction rates for $2\rho \leftrightarrow 3\rho$ scatterings:
\begin{align}
    \Gamma_{2\to3} = 2\times \frac{\text{number of } 2\to 3 \text{ collisions}}{N_\rho \Delta t} \; , \\
    \Gamma_{3\to2} = 3\times \frac{\text{number of } 3\to 2 \text{ collisions}}{N_\rho \Delta t} \; .
\end{align}
We show them in figure \ref{fig:  2 to 3 scattering rates} along with the analytic expectations for kinetic (but not chemical) equilibrium 
$\Gamma^\text{eq}_{2\to3}=n_\rho\,\langle \sigma v\rangle^\text{eq}_{2\to3}$ and $\Gamma^\text{eq}_{3\to2}=\frac{n_\rho^2}{2}\, \langle \sigma v^2\rangle^\text{eq}_{3\to2}$ with the thermally averaged cross sections as defined in equation \ref{eq: thermally averaged cross section definition}. To calculate these thermal averages we use the would-be chemical equilibrium temperature, i.e.\ the lower end of the grey band on the right side of figure~\ref{fig:  yield and temperature evolution numberchanging scattering}. 

For constant matrix elements the cross section $(\sigma v)_{2\to 3}$ is constant for relativistic particles. The time dependence of the interaction rate is therefore set only by the number density. In case of a constant comoving number density, one finds $\Gamma_{2\to3}\sim T^3$, so the rate decreases faster than the Hubble rate. This shows that
$2\rho \to 3\rho$ processes are active only if the combination of initial abundance produced through decays and matrix element is large enough to have $\Gamma_{2\to3}\gtrsim H$ immediately at production. In this case, the comoving number density increases further making number-changing processes even more important. This self-amplifying process only stops once the chemical equilibrium abundance is reached and the production of particles is balanced by the inverse $3 \to 2$ process. 

In the absence of further external particle injection, the number density is then once again covariantly conserved  (for relativistic particles), such that the number-changing rate drops faster than the Hubble rate and ultimately becomes inefficient. The out-of-equilibrium decay enhances the number density once more, potentially pushing the interaction rate again above the Hubble rate, such that also the high-energy peak thermalises.

For $3 \to 2$ scattering, we know from eq.~\eqref{eq: scattering prob 3 to 2} that the scattering rate scales as $\Gamma_{3\to 2}\sim \left|\mathcal{M}_{3\rho \to 2\rho} \right|^2 n_\rho^2/E^3$, where $E$ is the energy of the involved particles. For relativistic particles, we then find that $\Gamma_{3\to 2}\sim T^3$ in case of constant comoving number density. Since the equilibrium number density lies above the number density produced through the decays, the $3\to 2$ rate is initially small. As soon as the $2 \to 3$ scatterings increase the number density sufficiently, also $3 \to 2$ scatterings become important and start to balance the $2 \to 3$ scatterings. In case the matrix element is large enough, thermalisation is achieved and the interaction rate decreases again. The out-of-equilibrium decays and subsequent increase of $2\rho \to 3\rho$ scatterings also again increases the $3\rho \to 2\rho$ scattering rate for large enough matrix elements, which can lead to thermalisation also of the second peak. Deviations between the analytical and simulated rates reflect the fact that the phase-space distribution is non-thermal immediately after the production of the high-momentum peak because the considered rates are too small to instantly thermalise the injected particles.

To conclude this discussion, we point out that the final $\rho$ particle yield, corresponding to the predicted DM abundance, depends sensitively on the rate of number-changing processes. For example, the orange and purple lines correspond to matrix elements that differ only by a factor of 2, while the resulting abundances differ by an order of magnitude. Moreover, in the most decisive stage, when high-momentum particles are injected through out-of-equilibrium decays, there are visible differences between the actual interaction rates in the simulations and their analytical estimates assuming thermal distributions. This highlights the need for more sophisticated methods such as DSMC.

\subsection{Convergence}

As the DSMC stochastically solves the Boltzmann equation, each simulation inherently comes with a statistical error from the finite particle number. For example, the DM abundance predicted from our simulation is directly proportional to the total number $N_\text{end}$ of simulation particles at the end of the simulation, such that one can expect a relative error of the order of $1/\sqrt{N_\text{end}}$. For quantities obtained from the normalised phase-space distribution, such as the temperature $T_\rho$, the error can be estimated from the sample variance using the standard formula for Monte Carlo integration, which again predicts that the error decreases as $1/\sqrt{N_\text{end}}$. These error estimates can be directly confirmed by re-running the DSMC several times to obtain independent samples of the final distribution.

To check for convergence of our results with increasing particle number, we exemplarily performed three sets of simulations for $\left|\mathcal{M}_{3\rho \to 2\rho} \right|^2= 2\times 10^{-11}\, \textrm{GeV}$ for three different initial simulation volumes $V_i$, $4V_i$ and $9V_i$, corresponding to a final number of $\rho$ particles of $N_\text{end} = \mathcal{O}(45\,000),\, \mathcal{O}(180\,000),$ and $\mathcal{O}(405\,000)$, respectively. For each case we ran ten independent simulations and computed the mean and standard deviation for the final $\rho$ abundance of the simulation ensembles. 

\begin{figure}[tbp]
\centering
\includegraphics[width=\textwidth]{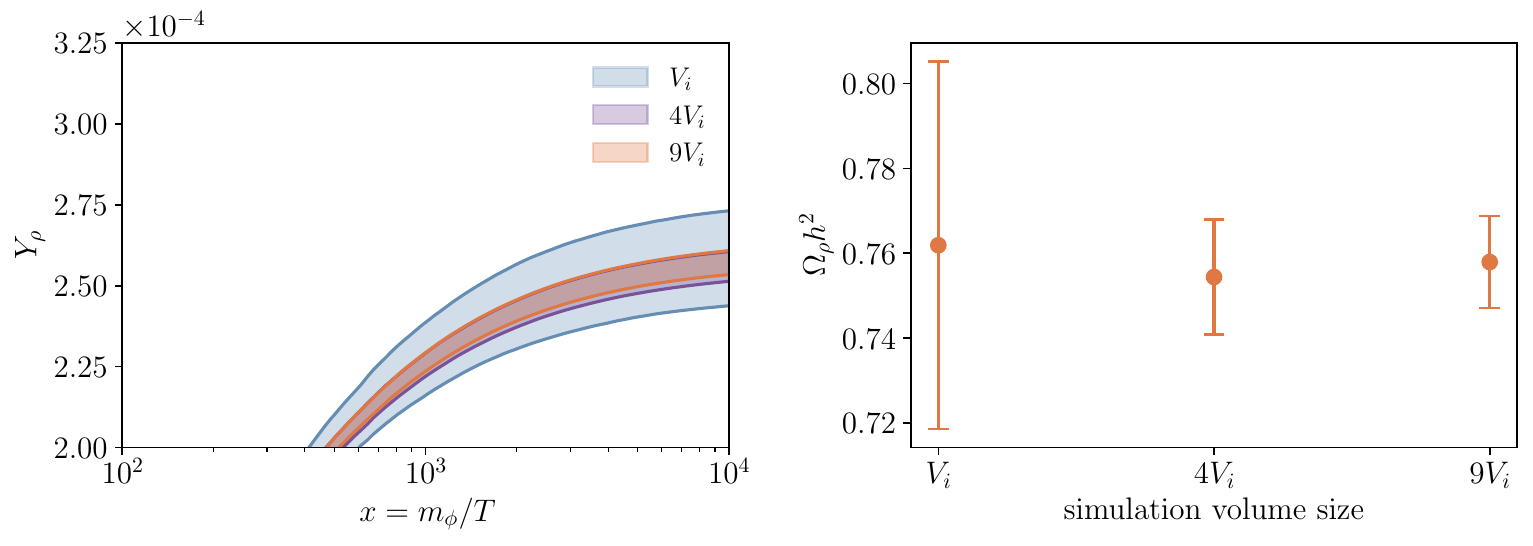}
\caption{Left: Evolution of $Y_\rho$ as a function of temperature for three different simulation sizes. The band shows the $\pm 1\sigma$ uncertainty obtained from ten independent simulation runs.  Right: Mean value and standard deviation of $Y_\rho$ for the different simulation sizes. The simulation volume of initial size $V_i$ corresponds to $\mathcal{O}(45000)$ $\rho$ particles at the end of the simulation.\label{fig: convergence check}}
\end{figure}

\begin{figure}[tbp]
\centering
\includegraphics[width=\textwidth]{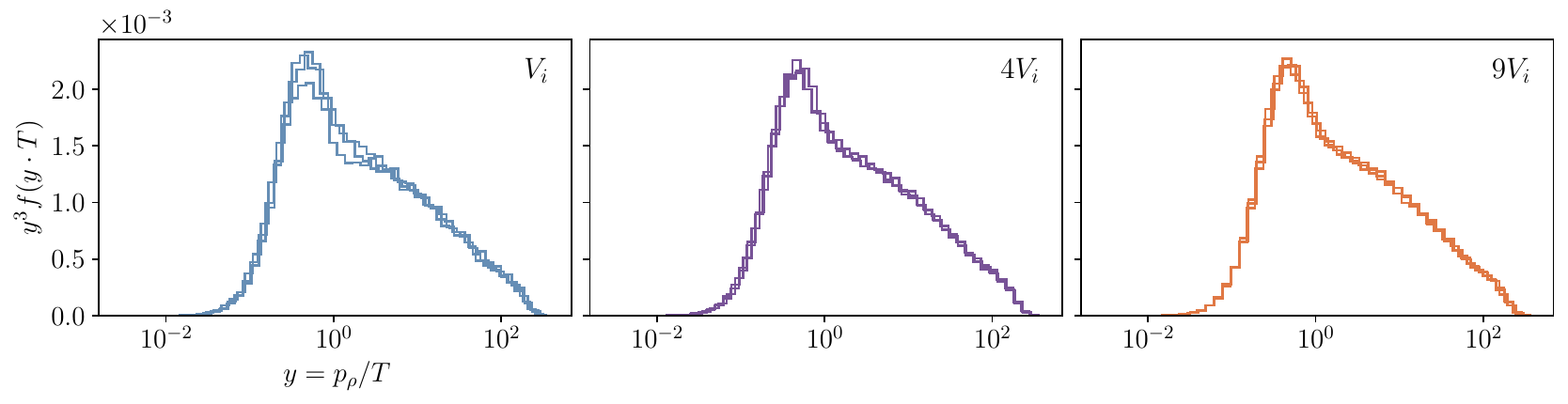}
\caption{Phase-space distribution after finished freeze-in for three different runs with a simulation volume of $V_i$ (left), $4V_i$ (middle) and $9V_i$ (right). \label{fig: convergence histograms}}
\end{figure}

In figure \ref{fig: convergence check}, we show on the left side the evolution of the $\pm 1\sigma$ uncertainty band for $Y_\rho$ obtained from the simulation ensembles and on the right side the ensemble average and standard deviation for $\Omega_\rho h^2$. We see that the results for the different volumes agree with each other and that the precision increases with simulation size. For the largest simulations, percent-level precision in the final abundance is achieved. We observe that the standard deviation roughly decreases as $1/\sqrt{N}$ with the number of simulation particles, as expected for a stochastic simulation. The actual uncertainty is larger than the naive expectation of a Poisson distribution with uncertainty $\sqrt{N}$, because all the number-changing processes (i.e.\ decays and $2\leftrightarrow 3$ scatterings) introduce correlations in the ensemble. In figure \ref{fig: convergence histograms}, we show the phase-space distribution after finished freeze-in for the different volumes for three different simulation runs. Convergence towards a unique phase-space distribution with increasing simulation volume is visible. We have also confirmed that quantities calculated from the phase-space distribution, such as the temperature, exhibit the expected convergence with increasing particle number.

\section{Conclusions and outlook}
\label{sec:conclusions}

The phase-space distribution of dark matter (DM) carries important information for cosmology but is challenging to calculate whenever thermal equilibrium cannot be assumed. Solving Boltzmann equations for DM production at the phase-space level hence becomes increasingly important in the era of precision cosmology. In this work, we showed that Monte Carlo simulations provide a highly useful tool in the context of non-equilibrium evolution. A particular advantage is that the relevant phase-space regions do not need to be known ahead of time, because the simulations can be used as an explorative tool to identify physically relevant momentum ranges. Within the framework of Direct Simulation Monte Carlo (DSMC), we studied freeze-out, decay, elastic as well as number-changing interactions for constant matrix elements, demonstrating that a wide range of DM production scenarios can be explored at the phase-space level.
We confirmed the correct implementation of the various reaction processes by cross-checking against analytical expectations.

To illustrate the relevant effects, we considered a toy model based on a well-known scenario with two-step production of DM through in-equilibrium and out-of-equilibrium decays of a parent particle. The resulting bimodal phase-space distribution is highly non-thermal, necessitating an unintegrated Boltzmann equation to study the effect of subsequent $2\to 2$ and $2\leftrightarrow 3$ scatterings. As expected, for large enough matrix elements, $2\to 2$ scattering leads to kinetic equilibrium, while $2 \to 3$ scattering also establishes chemical equilibrium. The DSMC approach is able to describe both the approach towards thermal equilibrium for large matrix elements as well as scenarios with smaller matrix elements, both of which cannot be modelled using approximations that assume thermal distributions.

A DSMC run predicts not only the final number of particles in the simulation volume corresponding to the DM relic abundance, but it also provides a random sample from the DM phase-space distribution. This sample can be used for efficient Monte Carlo integration of integrals involving the phase-space distribution function. These include simple expectation values such as the DM temperature, but also more complicated quantities such as the perturbations needed to calculate the suppression of small-scale structure in Boltzmann codes~\cite{Lesgourgues:2011rh}. For all these quantities, the statistical Monte Carlo error can be determined by running the simulation several times and reduced by increasing the simulation volume.

The present work did not aim to describe a realistic DM scenario. The toy model under consideration instead served as a proof of concept to show the flexibility and limitations of DSMC in context of DM production. So far, we have only considered constant matrix elements, but it is in principle straightforward to include more complicated matrix elements. In fact, in this context the DSMC approach can benefit strongly from established methods for Monte Carlo event generation, such as importance sampling or adaptive integration. However, in order to make the simulations efficient, tight upper bounds on the interaction cross section are needed for the matrix elements under consideration. Furthermore, we have not included quantum statistics effects, but these can be handled within DSMC as demonstrated in e.g.\ Ref.~\cite{Akita:2026xwu}.

An important challenge for the DSMC approach arises when considering scenarios where the particle number changes by orders of magnitude or when several processes of interest occur at vastly different rates. In such cases new strategies are needed to guarantee computational feasibility, based for example on splitting and merging algorithms that have been developed in the context of DSMC. In this context it will also be interesting to systematically benchmark the DSMC approach against alternative ways to solve unintegrated Boltzmann equations in order to identify which approach works best in which scenario. The present work demonstrates that it is worthwhile to explore these directions in the future in order to provide the community with new tools and techniques to understand the evolution of DM in the early universe and pin down its role in structure formation.

\acknowledgments

We thank Torsten Bringmann, Oleksii Ihnatenko and Thomas Schwetz for helpful discussions, Tim-Leon Klocke for valuable comments on the manuscript and Maksym Ovchynnikov for sharing the code from Ref.~\cite{Ovchynnikov:2024rfu}. 

\appendix
\section{Derivation of $2\leftrightarrow3$ scattering probability}
\label{app: 2 to 3}

In this appendix we provide additional details on the calculation of the scattering rates of number-changing processes. These have been derived in ref. \cite{Xu:2004mz}.

\paragraph{$2\rightarrow 3$ scattering. } 
The collision operator for $2\leftrightarrow 3$ scattering is given by
\begin{align}
\label{eq: collision 2 3}
\frac{\hat C_{23}}{E_1} =\;&
\frac{1}{2E_1}\frac{1}{2!}
\int \mathrm{d}\pi_2 \mathrm{d}\pi_3\frac{1}{2!}
\int \mathrm{d}\pi_1' \mathrm{d}\pi_2'
\, f_1' f_2'
\left|\mathcal{M}_{1'2' \to 123}\right|^2
\nonumber
(2\pi)^4\delta^{(4)}\!\left(p_1' + p_2' - p_1 - p_2 - p_3\right)
\nonumber\\
&+
\frac{1}{2E_1}
\int \mathrm{d}\pi_2 \frac{1}{3!}
\int \mathrm{d}\pi_1' \mathrm{d}\pi_2' \mathrm{d}\pi_3'
\, f_1' f_2' f_3'
\left|\mathcal{M}_{1'2'3' \to 12}\right|^2
\nonumber
(2\pi)^4\delta^{(4)}\!\left(p_1' + p_2' + p_3' - p_1 - p_2\right)
\nonumber\\
&-
\frac{1}{2E_1}
\frac{1}{2!}
\int \mathrm{d}\pi_2 \mathrm{d}\pi_3 \frac{1}{2!}
\int \mathrm{d}\pi_1' \mathrm{d}\pi_2'
\, f_1 f_2 f_3
\left|\mathcal{M}_{123 \to 1'2'}\right|^2
\nonumber 
(2\pi)^4\delta^{(4)}\!\left(p_1 + p_2 + p_3 - p_1' - p_2'\right)
\nonumber\\[1ex]
&-
\frac{1}{2E_1}
\int \mathrm{d}\pi_2 \frac{1}{3!}
\int \mathrm{d}\pi_1' \mathrm{d}\pi_2' \mathrm{d}\pi_3'\, f_1 f_2
\left|\mathcal{M}_{12 \to 1'2'3'}\right|^2
(2\pi)^4\delta^{(4)}\!\left(p_1 + p_2 - p_1' - p_2' - p_3'\right).
\end{align}
To obtain the probability of scattering in a time interval $\Delta t$ when choosing two initial particles in the momentum range $(\bf{p}_1, \bf{p}_1+\Delta p_1)$ and $(\bf{p_2}, \bf{p_2}+\Delta p_2)$ in a volume $\Delta x^3$, we proceed in the same way as for the $2\to 2$ scattering (by selecting the last term in eq.~\eqref{eq: collision 2 3} responsible for $2\rightarrow3$ scattering). The scattering probability is then given by 
\begin{align}
    P_{2\rightarrow3}=\frac{N_\textrm{coll}}{N_1 N_2}=\sigma_{2\rightarrow 3}v\frac{\Delta t}{\Delta x^3}
\end{align}
with 
\begin{align}
    \sigma_{2\rightarrow 3}v=\frac{1}{2E_1}\frac{1}{2E_2}\frac{1}{3!}\int \mathrm{d}\pi_1'\mathrm{d}\pi_2'\mathrm{d}\pi_2'\, \left| \mathcal{M}_{12 \to 1'2'3'} \right|^2 
(2\pi)^4
\delta^{(4)}\!\left(p_1 + p_2 - p_1' - p_2'-p_3'\right) \; .
\end{align}

\paragraph{$3\rightarrow 2$ scattering.}
To obtain the scattering probability, we again use the Boltzmann equation, now selecting the third term in eq.~\eqref{eq: collision 2 3}. 
The number of $3\rightarrow 2$ collisions during a time interval $\Delta t$ in a volume $\Delta x^3$ with $N_i$ particles in the momentum range $(\bf{p}_i, \bf{p}_i+\Delta p_i)$ is given by 
\begin{align}
    \frac{N_\textrm{coll}}{\Delta t\frac{1}{(2\pi)^3}\Delta x^3 \Delta p_1^3 }=& \frac{1}{2E_1}\frac{\Delta^3p_2}{(2\pi)^32E_2}\frac{\Delta^3p_3}{(2\pi)^32E_3}\frac{1}{2!} f_1f_2f_3 \nonumber \\
    & \times \int \mathrm{d} \pi_1' \mathrm{d}\pi_2' \left| \mathcal{M}_{123 \to 1'2'} \right|^2
(2\pi)^4
\delta^{(4)}\!\left(p_1 + p_2+p_3 - p_1' - p_2'\right)
\end{align}
The collision probability after selecting three particles with momenta in the range $(\bf{p}_i, \bf{p}_i+\Delta p_i)$ is then given by 
\begin{align}
    P_{3\rightarrow2}=\frac{N_\textrm{coll}}{N_1 N_2 N_3}=\frac{1}{8E_1E_2E_3}\frac{\Delta t}{(\Delta x^3)^2}I_{32}
\end{align}
where we defined the Lorentz-invariant integral expression
\begin{align}
    I_{32}=\frac{1}{2!}\int \mathrm{d} \pi_1' \mathrm{d}\pi_2'
\, \left| \mathcal{M}_{123 \to 1'2'} \right|^2
(2\pi)^4
\delta^{(4)}\!\left(p_1 + p_2 +p_3- p_1' - p_2'\right) \; .
\end{align}
Note that the additional factor of $1/2!$ in the collision operator appears due to initial state identical particles and is not included in the collision probability because we account for it by considering unordered triplets $N_\textrm{3-plets}= \binom{N}{3}$ in section \ref{sec: DM DSMC}.

\bibliographystyle{JHEP}
 \bibliography{biblio.bib}

\end{document}